\documentclass[11pt,preprint]{aastex}
\usepackage{epstopdf}
\usepackage{chngpage}
\usepackage{graphicx}
\usepackage{natbib}
\usepackage{mathrsfs}

\shorttitle{ELECTRON-POSITRON WIND}
\shortauthors{Geng et al.}

\begin{document}

\title{IMPRINTS OF ELECTRON-POSITRON WINDS ON THE MULTI-WAVELENGTH AFTERGLOWS OF GAMMA-RAY BURSTS}

\author{J. J. Geng\altaffilmark{1, 2}, X. F. Wu\altaffilmark{3, 4}, Y. F. Huang\altaffilmark{1, 2},
L. Li\altaffilmark{5, 6} and Z. G. Dai\altaffilmark{1, 2}}

\altaffiltext{1}{School of Astronomy and Space Science, Nanjing University, Nanjing 210046, China; hyf@nju.edu.cn}
\altaffiltext{2}{Key Laboratory of Modern Astronomy and Astrophysics (Nanjing University), Ministry of Education, China}
\altaffiltext{3}{Purple Mountain Observatory, Chinese Academy of Sciences, Nanjing 210008, China; xfwu@pmo.ac.cn}
\altaffiltext{4}{Joint Center for Particle Nuclear Physics and Cosmology of Purple Mountain Observatory-Nanjing University, Chinese Academy of Sciences, Nanjing 210008, China}
\altaffiltext{5}{Department of Physics, Stockholm University, AlbaNova, SE-106 91 Stockholm, Sweden}
\altaffiltext{6}{Erasmus Mundus Joint Doctorate in Relativistic Astrophysics}

\begin{abstract}
Optical re-brightenings in the afterglows of some gamma-ray bursts (GRBs) are
unexpected within the framework of the simple external shock model.
While it has been suggested that the central engines of some GRBs are newly born magnetars,
we aim to relate the behaviors of magnetars to the optical re-brightenings.
A newly born magnetar will lose its rotational energy in the form of Poynting-flux,
which may be converted into a wind of electron-positron pairs through some magnetic dissipation processes.
As proposed by \cite{Dai04},
this wind will catch up with the GRB outflow and a long-lasting reverse shock would form.
By applying this scenario to GRB afterglows,
we find that the reverse shock propagating back into the electron-positron
wind can lead to an observable optical re-brightening
and a simultaneous X-ray plateau (or X-ray shallow decay).
In our study, we select four GRBs, i.e., GRB 080413B, GRB 090426, GRB 091029, and GRB 100814A,
of which the optical afterglows are well observed
and show clear re-brightenings. We find that they can be well interpreted.
In our scenario, the spin-down timescale of the magnetar should
be slightly smaller than the peak time of the re-brightening, which can provide
a clue to the characteristics of the magnetar.
\end{abstract}


\section{INTRODUCTION}
\label{sect:intro}

Recent multi-wavelength observations of afterglows of gamma-ray bursts (GRBs) have revealed
some puzzling features \citep{Panaitescu06,Panaitescu11,Li12,Liang13},
impelling the researches on the afterglow theory.
X-ray observations from {\it Swift}/XRT \citep{Gehrels04,Burrows05} have resulted in diverse light curves,
but a canonical light curve consisting of a steep decay, a plateau, and a normal decay phase has been suggested \citep{Zhang06}.
The plateau phase can not be explained in the framework of the simple standard external shock
scenario, i.e. synchrotron radiation produced by the interaction of a relativistic ejecta with
the circum-burst medium (see \citealt{Piran93,Meszaros97,Sari99} for reviews).
On the other hand, some optical afterglows show re-brightenings at late stages ($\sim 10^4 - 10^5$ s), together with
a bump occurring in the X-ray band in some cases (e.g., GRB 120326A, \citealt{Melandri14,Hou14,Laskar15}),
or without clear counterpart features in X-rays (e.g., GRB 100814A, \citealt{Pasquale15};
GRB 081029, \citealt{Nardini11}; GRB 100621A, \citealt{Greiner13}).
These re-brightenings also deviate from the expectations of the external
shock model, which calls for a refined afterglow model.

Several scenarios have been proposed to explain these unexpected features.
Some authors invoke energy injection processes \citep{Dai98a,Dai98b,Fan06b,vanEerten14}
or forward shocks (FS) refreshed by late shells \citep{Zhang02} to produce the X-ray plateau.
Meanwhile, circum-burst density enhancements \citep{Dai03,Lazzati02},
two-component jets \citep{Berger03,Huang04}, or varying microphysical parameters \citep{Kong10}
have been invoked to explain the optical re-brightenings.
However, recent studies indicate the density enhancements model may not work for
significant re-brightenings \citep{vanEerten09,Gat13,Geng14}.
If one wants to interpret a large sample of afterglows, which model is preferred is still uncertain.
Although there are various candidate models for afterglows,
the possible central engines of GRBs are mainly limited in two kinds of objects.
It is generally believed that the central engines of GRBs
should either be rapidly spinning, strongly magnetized neutron stars (magnetar, \citealt{Dai98a,Dai98b,Heger03,Dai06,Metzger11,Lv14}) 
or black holes \citep{Woosley93,MacFadyen99,Lei13}. This indicates that there is probably a common origin 
for the optical re-brightenings and the simultaneous X-ray features of a group of afterglows.
Therefore, it is reasonable to develop a model which, after taking into account the physics of the central engine,
can explain the multi-wavelength afterglows simultaneously.

From the visual perspective, the optical re-brightening seems to be a new component emerging
when the flux from the FS gets dim. In the two-component jet model,
this kind of late-time component naturally corresponds to a wide jet. However, it may
have difficulties in explaining the chromatic evolution of the afterglows in some cases \citep{Pasquale15}.
The new component can also be generated in another model, i.e., a long-lasting reverse shock (RS) can play the role.
If the FS is re-energized by a late outflow (e.g., an electron-positron-pair wind, \citealt{Dai04,Yu07a,Yu07b}),
then an RS will form behind the FS. As the RS propagating into the outflow, it
would contribute to the emission, and its flux can even exceed that from the FS.
Moreover, the emission from the RS can help to make diverse light curves in principle
since the microphysical parameters may differ from those of the FS.
Therefore, it is reasonable to argue that some light curves with re-brightenings
are due to the existence of a long-lasting RS.
On the other hand, the long-lasting RS should be common since it is
a natural consequence of the energy injection. In previous studies,
it has been proposed that continuous energy injections in the form of  pure
Poynting-flux coming from the central engines may play a key role
in explaining some special light curves, through its effects on the evolution of the FS
 \citep{Zhang06,Liu09,Geng13,Yu15}.
However, in such a scenario without an RS, the emission from the FS could not
account for the new components of the re-brightenings discussed here.

In this paper, we propose the ultrarelativistic electron-positron-pair
($e^+e^-$) wind model to interpret the afterglows with special re-brightenings.
After a GRB,  the remaining object of the progenitor
may be a magnetar, which should lose its rotational energy by ejecting
a continuous Poynting-flux. We assume that the Poynting-flux is
converted into an ultrarelativistic $e^+e^-$ wind beyond
a particular radius ($\sim 10^{15}$ cm).
When the $e^+e^-$ wind catches up with the FS, an RS will form and
propagate back into the $e^+e^-$ wind. The wind will be shocked
and heated by the RS, and radiation from these
$e^+e^-$ can account for the optical re-brightenings at late time.
It is worthwhile to note that this scenario was originally proposed by \cite{Dai04} and then
used to explain the X-ray plateau and bump by \cite{Yu07a} and \cite{Dai12}.
However, the method used to solve the shock dynamics in the current work is different
and we aim to explain optical re-brightenings. The mechanism of converting
the Poynting-flux into ultrarelativistic $e^+e^-$ wind is still unknown.
Previous researches indicate that magnetic dissipation may help to
explain why pulsar wind nebulae are powered by particle-dominated winds
\citep{Rees74,Lyubarsky01,Porth13,Metzger14}.
Such a dissipation may also occur in millisecond magnetars.

In our study, we select as examples four GRBs, of which high quality multi-wavelength observational
data are available. They are all characterized by optical re-brightenings but without
corresponding behaviors being detected in X-rays. We show that they can be well explained in our model.
Our paper is organized as follows. In Section 2, we briefly describe the dynamic methods
used in our work and the formulae for calculating the radiation.
In Section 3, we show how our scenario would work to explain
the re-brightenings. The results for each GRB are presented separately in Section 4.
The conclusions are summarized in Section 5.

\section{HYDRODYNAMICS AND RADIATION}
When a relativistic outflow propagates into the circum-burst medium,
an FS will form. If the central engine is long-lived and a continuous
wind is ejected, an RS is expected to propagate back into the wind.
The dynamics of such an FS-RS system can be numerically solved by
considering energy conservation as proposed by \cite{Geng14},
or through the mechanical consideration as done by Beloborodov \&
Uhm (2006; also see Uhm 2011).
For clarity, these two methods are described separately in the Appendixes
A.1 and A.2.

Now, we show that the dynamical results of these two methods
are consistent with each other within our $e^+e^-$ wind scenario.
The Poynting-flux luminosity $L_{\rm w} (t_{\rm obs})$ of a newly born magnetar can be
derived from the magnetic dipole radiation \citep{Shapiro83}, i.e.,
\begin{equation}
L_{\rm w} \simeq 4.0 \times 10^{47} B_{\rm NS,14}^2 R_{\rm NS,6}^6 P_{{\rm NS},-3}^{-4} \left(1+\frac{t_{\rm obs}}{T_{\rm sd}}\right)^{-2} \rm{erg}~\rm{s}^{-1},
\end{equation}
where $B_{\rm NS}$ is the strength of the surface magnetic field of the magnetar,
$R_{\rm NS}$ is the radius and $P_{\rm NS}$ is its spin period.
$T_{\rm sd} \simeq 5.0 \times 10^4 (1+z) B_{\rm NS,14}^{-2} I_{45} R_{\rm NS,6}^{-6} P_{\rm NS,-3}^2$ s
is the spin-down timescale,
where $I$ is the moment of inertia of the magnetar.
The convention $Q_x = Q/10^x$ in cgs units is adopted hereafter.
For simplicity, we fix $I = 10^{45}~{\rm g}~{\rm cm}^2$ , $R_{\rm NS} = 10^6$~cm
and $P_{\rm NS} = 1$~ms throughout this work, then $L_{\rm w}$ is mainly determined
by only one parameter, i.e., $B_{\rm NS}$.
We assume the Poynting-flux is converted into an $e^+e^-$ pair wind beyond $10^{15}$ cm,
which is the typical deceleration radius of the outflow.
The particle density in the comoving frame of the unshocked wind (or called Region 4)
at the radius $r$ is then
\begin{equation}
n_4^{\prime} = \frac{L_{\rm w}}{4 \pi r^2 \Gamma_4^2 m_e c^3},
\end{equation}
where $m_e$ is the mass of electron, $c$ is the speed of light
and $\Gamma_4$ is the Lorentz factor of the unshocked wind.
Note that the FS-RS system can be divided into four regions (see Appendix A.1),
the quantities of Region ``$i$'' are denoted by subscripts ``$i$''.
In this paper, the superscript prime ($\prime$) is used to denote the quantities in the shock
comoving frame while characters without prime denote quantities in the
observer frame.

For an outflow with an isotropic kinetic energy of
$E_{K,\rm{iso}} = 1.0 \times 10^{53}$ erg and
an initial Lorentz factor of $\Gamma_0 = 300$,
we set the number density of the ambient medium $n_1$ is 1~cm$^{-3}$,
$\Gamma_4$ is $10^4$, and $B_{\rm NS}$ is $2 \times 10^{14}$ G.
Then the evolution of the bulk Lorentz factor of the FS ($\Gamma_2$)
and the RS ($\Gamma_3$) can be obtained using the two methods mentioned above.
Figure 1 shows the results from the energy conservation method
and the mechanical method. We see that the difference between the two
solutions is tiny, which means they are consistent with each other.
The two solutions approach the solution given by \cite{Blandford76}
(referred to as the BM solution below) at late times,
which is foreseeable since $L_{\rm w}$ finally becomes very weak and
the role of the RS is insignificant.
In our following study, we will adopt the mechanical method
to solve the dynamics of the FS-RS system.

To calculate the afterglow light curves of the FS-RS system, we consider
both the synchrotron radiation and the inverse compton radiation from shock-accelerated
electrons. Detailed formulae are described in Appendix B.

\section{APPLICATION}
In our study, we have collected the afterglow data of four GRBs, i.e.,
GRB 080413B \citep{Filgas11}, GRB 090426 \citep{Nicuesa11},
GRB 091029 \citep{Filgas12} and GRB 100814A \citep{Pasquale15}.
The observational data are of high quality, and all the four events
show clear optical re-brightenings at $\sim 10^4 - 10^5$ s.
The rising of the re-brightenings is generally smooth and shallow,
contrary to many other GRBs which brightens sharply \citep{Geng13}.
This indicates the re-brightening component is a result of
a continuous behavior, which can be well achieved in our model.
A brief summary of the observations of these GRBs is presented in Table 1.
We will interpret these afterglows by using our model
and discuss the underlying relation between the afterglow behavior and
the central engine.

Our sample of afterglows are characterized by a double-bump structure.
For the early onset bump (it may not show up due to the lack of early observations),
we suggest that it is connected with the typical synchrotron frequency
crossing the optical band, or the FS being in the coasting phase.
The second re-brightening bump could be due to the emerging RS component at about $T_{\rm sd}$.
We take GRB 100814A as an example to show how this double-bump
structure can be consistently reproduced in our model and
how the physical parameters can be estimated from analytical derivations.
Below, we use the convention $F_{\nu,i} \sim t_{\rm obs}^{-\alpha} \nu^{-\beta}$,
where $F_{\nu,i}$ is the flux density of Region $i$, $\alpha$ and $\beta$
are the power-law indices.

The first bump appearing in the optical afterglows of GRB 100814A is
roughly at $T_{\rm peak,1} \sim 300$ s and the rising temporal index is $\alpha_{\rm rise} = -2.5$.
In the framework of the external shock model,
it indicates that the FS is in the coasting phase \citep{XueRR09}.
We assume the isotropic kinetic energy of the outflow is $E_{K,{\rm iso}}$,
then the initial Lorentz factor $\Gamma_0$ can be calculated as
$\Gamma_0 = \left[\frac{17 E_{K,{\rm iso}} (1+z)^3}{64 \pi n_1 m_p c^5 T_{\rm peak,1}^3}\right]^{1/8}$.
Assuming $E_{K,{\rm iso}}$ equals the energy released in the prompt phase ($E_{\gamma,{\rm iso}} = 7 \times 10^{52}$ erg)
and using typical value of $n_1 = 0.1$ cm$^{-3}$, we obtain $\Gamma_0 \simeq 150$.
At the end of the coasting phase, $\nu_{m,2}$ should have crossed the optical
band since the observed optical flux decreases after $T_{\rm peak,1}$.
This requires $\nu_{m,2}(T_{\rm peak,1}) \leq 10^{15}$ Hz, i.e.,
\begin{equation}
\frac{1}{1+z} \left(\frac{p_2-2}{p_2-1}\right)^2 \epsilon_{e,2,-1}^{2} \epsilon_{B,2,-1}^{1/2}
n_{1,0}^{1/2} \leq 3.5 \times 10^{-5},
\end{equation}
where $z$ is the redshift of the GRB, $p_i$ is the power-law index of the distribution of electrons
in Region $i$, $\epsilon_{e,i}$ is the fraction of the total energy carried by
the electrons, and $\epsilon_{B,i}$ is the ratio of the magnetic field energy to the total energy.

After the coasting phase, we assume the evolution of $\Gamma_2$ obeys $\Gamma_2 = A R^{-g}$.
According to Section 2, we can still use the  BM
solution, i.e., $A = \left[\frac{17 E_{K, {\rm iso}}}{8 \pi n_1 m_p c^2}\right]^{1/2}$,
and $g = \frac{3}{2}$.
Then the typical synchrotron frequency of Region 2 is given by
\begin{equation}
\nu_{m,2} = 5.5 \times 10^{13} \left(\frac{1+z}{2}\right)^{1/2} \left(\frac{p_2-2}{p_2-1}\right)^2
\epsilon_{e,2,-1}^2 \epsilon_{B,2,-1}^{1/2}
E_{K,{\rm iso},53}^{1/2} t_{\rm obs, day}^{-3/2}~\rm{Hz},
\end{equation}
and the peak flux density at $D_L$ (luminosity distance) is
\begin{equation}
F_{\nu,{\rm max},2} = 8.6 \times 10^{2} \left(\frac{1+z}{2}\right)
\epsilon_{B,2,-1}^{1/2} E_{K,{\rm iso},53} n_{1,0}^{1/2} D_{L,28}^{-2}
~\rm{mJy}.
\end{equation}
For Region 3, $L_{\rm w}$ can be taken as a constant within $T_{\rm sd}$,
and the relative Lorentz factor $\Gamma_{43} \simeq \Gamma_4/(2\Gamma_2)$,
then the typical synchrotron frequency is given by
\begin{equation}
\nu_{m,3} = 6.9 \times 10^{10} \left(\frac{1+z}{2}\right)^{-3/2} \left(\frac{p_3-2}{p_3-1}\right)^2
L_{{\rm w},47}^{1/2} \Gamma_{4,4}^{2} \epsilon_{e,3,-1}^2 \epsilon_{B,3,-1}^{1/2}
E_{K,{\rm iso},53}^{-1/2} n_{1,0}^{1/2} t_{\rm obs, day}^{1/2}~\rm{Hz}.
\end{equation}
The cooling frequency is
\begin{equation}
\nu_{c,3} = 1.4 \times 10^{16} \left(\frac{1+z}{2}\right) L_{{\rm w},47}^{-3/2} \epsilon_{B,3,-1}^{-3/2}
E_{K,{\rm iso},53} n_{1,0}^{-1} t_{\rm obs, day}^{-2}~\rm{Hz},
\end{equation}
and the peak flux density is
\begin{equation}
F_{\nu,{\rm max},3} = 27.4 \left(\frac{1+z}{2}\right)^{1/4}
L_{{\rm w},47}^{3/2} \Gamma_{4,4}^{-1}
\epsilon_{B,3,-1}^{1/2} E_{K,{\rm iso},53}^{-1/4} n_{1,0}^{1/4} t_{\rm obs, day}^{3/4}
D_{L,28}^{-2}~\rm{mJy}.
\end{equation}
According to Equations (6-7),  the condition of $\nu_{m,3} < \nu_{c,3}$
is usually valid within $\sim 10^5$ s.

After the first peak, the optical afterglow enters the slow decay phase,
during which the temporal index is $\alpha_{\rm decay} \simeq 0.72$.
On the other hand, $\alpha_{\rm decay}$ is predicted as $\frac{3(p_2-1)}{4}$
when $\nu_{m,2} < \nu_{\rm opt} < \nu_{c,2}$, which gives $p_2 \simeq 2.0$.

Just before the emergence of the RS component (i.e., $F_{\nu,3}$), $F_{\nu,2}$ should
equal $F_{\nu,3}$ at a certain time. We denote this time as $T_{\rm dent}$.
For the regime $\nu_{m,i} < \nu < \nu_{c,i}$, we have
$F_{\nu,i} = F_{\nu,{\rm max},i} \left(\frac{\nu}{\nu_{m,i}}\right)^{-(p_i-1)/2}$.
The $g$ band data shows $T_{\rm dent} \simeq 1.2 \times 10^4$ s,
and the flux density at $T_{\rm dent}$ is $\simeq 7 \times 10^{-2}$ mJy.
Therefore we can get $F_{\nu,2} (T_{\rm dent}) = F_{\nu,3} (T_{\rm dent}) = 3.5 \times 10^{-2}$ mJy,
which further gives
\begin{eqnarray}
\left(\frac{1+z}{2}\right)^{(p_2+3)/4}
\left(\frac{p_2-2}{p_2-1}\right)^{p_2-1} \epsilon_{e,2,-1}^{p_2-1} \epsilon_{B,2,-1}^{(p_2+1)/4} E_{K,\rm{iso},53}^{(p_2+3)/4} n_{1,0}^{1/2} D_{L,28}^{-2} \nonumber \\
= 4.1 \times 10^{-5} \left(\frac{\nu_{\rm opt}}{1.1 \times 10^{15}~{\rm Hz}} \right)^{(p_2-1)/2},
\end{eqnarray}
and
\begin{eqnarray}
\left(\frac{1+z}{2}\right)^{(-3 p_3+4)/4}
\left(\frac{p_3-2}{p_3-1}\right)^{p_3-1}
\epsilon_{e,3,-1}^{p_3-1} \epsilon_{B,3,-1}^{(p_3+1)/4}
E_{K,\rm{iso},53}^{-p_3/4} n_{1,0}^{p_3/4} L_{\rm{w},47}^{(p_3+5)/4} \Gamma_{4,4}^{p_3-2} D_{L,28}^{-2} \nonumber \\
= 5.6 \times 10^{-3} \left(\frac{\nu_{\rm opt}}{2.6 \times 10^{10}~{\rm Hz}} \right)^{(p_3-1)/2},
\end{eqnarray}
where $\nu_{\rm opt} = 6.4 \times 10^{14}$ Hz.

The flux from the RS should reach its peak at $T_{\rm sd}$
since the flux decays when $t_{\rm obs} > T_{\rm sd}$ (see below).
In principle we can set $T_{\rm sd}$ as the peak time of the
second bump, e.g, $T_{\rm sd} = T_{\rm peak,2}$.
However,  the effect of the equal arrival time surface (EATS) would
delay the peak time in the optical bands. On the other side, the peak time in X-rays
will be less affected. In the case of GRB 100814A, we notice that
there may be a small structure in the X-ray afterglow at $\sim 4 \times 10^4$ s,
which gives $T_{\rm sd} \simeq 4 \times 10^4$ s.
We take $I = 10^{45}~{\rm g}~{\rm cm}^2$ , $R_{\rm NS} = 10^6$~cm
and $P_{\rm NS} = 1$~ms as typical values, then
$B_{\rm NS}$ can be constrained from $T_{\rm sd}$, and
$L_{\rm w}$ can be obtained from Equation (1).

When $T_{\rm dent} < t_{\rm obs} < T_{\rm peak,2}$,
$L_{\rm w}$ cannot be treated as constant since $t_{\rm obs}$
is approaching $T_{\rm sd}$.
In this case, we have
$\nu_{m,3} \propto L_{\rm w}^{1/2} t_{\rm obs}^{1/2}$,
$F_{\nu,{\rm max},3} \propto L_{\rm w}^{3/2} t_{\rm obs}^{3/4}$,
and
$F_{\nu,3} \propto \nu_{m,3}^{(p_3-1)/2} F_{\nu,{\rm max},3} \propto
L_{\rm w}^{(p_3+5)/4} t_{\rm obs}^{(p_3+2)/4}$.
A power-law fitting to this segment gives the observed temporal index
as $\simeq -0.5$. If we assume the evolution of $L_{\rm w}$
during this segment is roughly $\propto t_{\rm obs}^{-0.3}$,
then $p_3$ can be inferred to be 2.4.
After the second peak ($t_{\rm obs} > T_{\rm peak, 2}$),
since $L_{\rm w}$ evolves as $\propto t_{\rm obs}^{-2}$,
$\nu_{m,3}$ will decay with time and $\nu_{c,3}$ will increase.
As a result, none of them will cross the optical band,
i.e. $\nu_{m,3}<\nu_{\rm opt}<\nu_{c,3}$.
The optical spectral index at this time becomes $\sim 0.7$,
which further gives $p_3 = 2.4$ since $(p_3-1)/2 = 0.7$.
This value is consistent with that obtained before.

The jet break time $t_{j}$ is hard to determine since the FS flux is always lower than
that of the RS component after $T_{\rm dent}$.
In our fitting, we assume $t_{j} \simeq 10^5$ s, which requires
the half-opening angle of the jet $\theta_j \sim 0.058$ rad \citep{Lu12}.
We have little knowledge on the bulk Lorentz factor of the $e^+e^-$ wind, $\Gamma_4$.
However, a value ranging from $10^4$ to $10^6$ may be reasonable \citep{Yu07b}.
For GRB 100814A, we take $\Gamma_4 \simeq 10^5$.
There are still four free parameters left, i.e., $\epsilon_{e,2}$,
$\epsilon_{B,2}$, $\epsilon_{e,3}$ and $\epsilon_{B,3}$.
Using conditions given in Equations (3), (9-10),
and an underlying relation $\epsilon_{e,3} = 1-\epsilon_{B,3}$,
we get $\epsilon_{e,2} \simeq 5.5 \times 10^{-2}$, $\epsilon_{B,2} \simeq 1.2 \times 10^{-3}$,
and $\epsilon_{B,3} \simeq 0.02$.
In our subsequent study, we use these parameter values as
a guide to perform numerical calculations, adjusting the parameters slightly to get
a visually good agreement to the observed afterglow light curves (see Figure 2).
The final values (see in Table 2) of the parameters do not deviate significantly from the above derivations.
The high-energy $\gamma$-ray afterglow of GRB 100814A is also calculated.
Figure 3 presents the flux density at 100 MeV, in which
the SSC flux from the RS leads to a small bump in the light curve at $\sim 10^5$ s.
This is a special feature predicted by our model.
High-energy observations in the future can help to test the prediction.
We apply the above procedure to the afterglows of other
three GRBs (see Figures 4-6). The derived parameters
are also listed in Table 2.

In order to provide a meaningful constraint on the parameters,
we make use of the data fitting code named emcee \citep{Foreman13},
which is a widely used tool based on the Markov chain Monte Carlo (MCMC) simulations.
There are totally 11 parameters for each GRB.
It would be difficult to solve the problem if we set all these parameters
as free parameters.
The degeneracy between the parameters is one
major reason. Even for 7 parameters related to the FS,
this degeneracy may be significant \citep{Ryan13,vanEerten15}.
The other reason is the incredible computational cost in the MCMC sampling.
Since we are mostly interested in the parameters connected with the RS,
we finally chose  $B_{\rm NS}$, $\Gamma_4$ and $\epsilon_{B,3}$ as free parameters.
The allowed ranges of these parameters in the Monte Carlo simulation is set to be
[0.1,100]$\times 10^{14}$ G, [0.1,100] and [$1.0 \times 10^{-6}$,0.9] respectively.
Other parameters are fixed by using the values of Table 2.
The simulation results are presented in Table 3.
Figure 7 shows the corner plot of the fit for GRB 080413B as an example,
which consists of the marginalized distributions of each parameter and
the covariances between pairs of parameters.

The sensitivity of the results (in Table 3) on fixing other parameters can be roughly analysed.
In order to approach a good fitting to a specific light curve with re-brightening,
the early afterglow is attributed to the FS and the re-brightening is attributed to the RS.
In the case of small $B_{\rm NS}$ (the re-brightening occurs relatively late),
the dynamics of the FS is slightly affected by the $e^+e^-$ wind.
That would make some FS parameters independent of the RS parameters.
Taking GRB 080413B as an example, the early peak of the afterglow indicates $\nu_{m,2}$
should be crossing the optical band at $\sim 10^2$ s (Figure 4).
If one applies the MCMC fitting with the FS parameters,
according to Equations (4) and (5), one would find $\epsilon_{B,2}$ and $E_{K,{\rm iso}}$ are
degenerate in some extent, and they are in negative correlation with
$\epsilon_{e,2}$ or $n_1$ (also see Figure 1 in \citealt{Ryan13}).
On the other hand, according to Equation (10), $E_{K,{\rm iso}}$ and $n_1$
would be involved in calculations of the flux from the RS.
Both of them are expected to be negatively correlated with $\Gamma_4$ in the corner plot.
To judge whether the results in Table 3 are meaningful, we can do the following experiment.
Now, we fix the parameters in Table 3, and leave parameters $\epsilon_{e,2}$,
$\epsilon_{B,2}$, $n_1$ free. The MCMC fitting (see Figure 8) gives $\epsilon_{e,2} = 0.043^{+0.013}_{-0.009}$,
$\epsilon_{B,2} = 0.027^{+0.011}_{-0.008}$, $n_1 = 0.12^{+0.005}_{-0.006}$ respectively.
These values do not deviate much from those in Table 2.
In Figure 8, we can see $n_1$ is well constrained, which implies the results in Table 3 are
sensitive to $n_1$. This could be understandable since $n_1$ is involved in
the calculation of the dynamics of the RS and other parameters of the RS are fixed now.
This feature implies that the results in Table 3 would become robust when
the early observational data (mainly related with the FS) could constrain
parameters $E_{K,{\rm iso}}$ and $n_1$ well.

\section{DISCUSSIONS}

The four GRBs studied here actually have been investigated by other authors
in the literature. Previous efforts failed to explain many of the
puzzling behaviors of these events. Here, we would like to discuss
how our mechanism can improve the modeling.

\subsection{GRB 080413B}
The afterglow  of  GRB 080413B has been explained by the on-axis
two-component jet model \citep{Filgas11}. In this model, the narrow ultra-relativistic
jet is responsible for the initial decay, while the moderately relativistic wide jet is
expected to  account for the late re-brightening. The collapsar model of
long-duration GRBs offers a possible
mechanism for generating a two-component jet \citep{Ramirez02,Kumar15}.
According to \cite{Filgas11}, the power-law index of the narrow jet electrons
can be derived as $p  \simeq 1.44$ ($\nu_m < \nu_{\rm opt} < \nu_c$) from
the observed optical temporal index  of $\alpha \simeq 0.73$.
However, such an electron distribution with $p$ significantly less than 2 seems to be too hard.
Simulation on Fermi acceleration of charged particles by relativistic shocks indicates $p$ is
$\simeq 2.26 \pm 0.04$ in the ultrarelativistic limit $\Gamma \gg 1$ \citep{Lemoine03}.
The investigation on the distribution of $p$ from {\it Swift} GRB afterglows
supports a Gaussian distribution centered at $p = 2.36$ with a width of 0.59 \citep{Curran10}.

In our model, this difficulty no longer exists. The evolution of the Lorentz factor of FS
before $T_{\rm sd}$ deviates from the BM solution
($\Gamma \propto R^{-g}, g = \frac{3}{2}$) due to the effect of the
$e^+e^-$ wind, i.e., $g$ should be $< \frac{3}{2}$. Using the closure
relation of $0.73 = \frac{4 g}{2g+1} \frac{p_2-1}{2}$, we can obtain $p_2= 2.1$
if $g \simeq 1 $. Thus we can fit the light curves with a reasonable $p$.
Additionally, in the two-component jet model, the optical spectral index (pre-jet break) of
$\alpha \simeq 0.9$ indicates $p = 1.8$ ($\nu_{\rm opt} > \nu_c > \nu_m$) for the wide jet.
In our model, $\nu_c$ is greater than $\nu_{\rm opt}$ but slightly smaller than $\nu_{\rm X}$,
which gives $p_3 = 2.8$ (see in Table 2). This value again is more acceptable.

\subsection{GRB 090426}
GRB 090426 is a short GRB according to its duration (Nicuesa Guelbenzu et al. 2011).
However, some authors argued that GRB 090426 may be connected with a collapsar
event, rather than the merger of two compact objects \citep{Antonelli09,Xin11,Levesque10}.
In this case, a magnetar could be naturally involved.
The high density of $n_1 = 50$ cm$^{-3}$ derived from our fitting is consistent with
the lower limit of 10 cm$^{-3}$ given by \cite{Xin11}.
The optical and X-ray light curves of GRB 090426 could be fitted by
the two-component jet model of \cite{Nicuesa11}. But we notice that the first
break (around 300 s) in the $R_{\rm c}$ band light curve cannot be explained by the
jet break of the narrow-jet component \citep{Nicuesa11}. On the contrary, it
can be reasonably explained as the cessation of the energy
injection episode \citep{Xin11}. This is a natural consequence of our model.
We have shown that the dynamics of the FS is significantly affected by the $e^+e^-$ wind
at early times, but it should be in accord with the BM solution at late stages.
In Figure 5, we can see that this break corresponds to the turn over of the
evolution of $\nu_{c,\rm{FS}}$.
According to \cite{Nicuesa11}, the steep decay of the optical afterglow after $3 \times 10^4$ s
is due to the sideways expansion of the wide jet. However, the sideways expansion
of a relativistic jet may be a very slow process \citep{Zhang09}, which disfavors
the short interval between the peak time ($\sim 10^4$ s) and
the jet break time in GRB 090426.
In our scenario, this steep decay is a natural result of the quick decline of
$L_{\rm w}$ after the spin-down timescale.

\subsection{GRB 091029}
The optical and X-ray light curves of the afterglow of GRB 091029 are not similar to each other
and are hard to be explained by a simple model.
It is very peculiar that the optical spectral index decreases between 0.4 ks and
9 ks, and increases later, but at the same time, the X-ray spectral index
is almost a constant.
\cite{Filgas12} examined several scenarios and argued that
a two-component jet can basically explain the observations.
The hardening of the optical spectrum can be explained by assuming that
the electron power-law index changes with time.
In our scenario, the hardening of the optical spectrum before $10^4$ s is
also due to the varying electron index $p$ of the FS \citep{Filgas12,Kong10}.
But after $10^4$ s, the softening of the optical spectrum is caused by the rising flux from the RS.
According to \cite{Filgas12}, the X-ray spectral index is $\simeq 1.1$ over the entire time window.
In our model, the X-ray band is in the $\nu > \nu_c > \nu_m$ regime at $10^3$ s,
which leads to $p_2 = 2.2$. Note that the optical flux from the RS should be
dominated after $10^4$ s, $p_3$ should be relatively larger
in order to match the softening spectrum.
In our numerical calculations, we finally takes $p_2 = 2.2$, $p_3 = 2.3$.
The broad-band afterglow of GRB 091029 can be fitted quite well.

\subsection{GRB 100814A}
Possible interpretations for the afterglow of GRB 100814A have been discussed
by \cite{Pasquale15}.
In \cite{Pasquale15}, the optical re-brightenings are attributed to the FS, when $\nu_{m}$ of the FS
is crossing the optical band, and the RS is generated by the late shells that
collide with the trailing ones.
In our scenario, the re-brightening is due to the emerging of the RS flux,
while the RS forms due to the injection of continuous $e^+e^-$ winds.
Our numerical results indicate that the optical flux from the FS peaks at around
400 s, when $\nu_{m}$ of the FS crosses the optical band (see the lower panel of Figure 2).
The observed $R_c$ band light curve (see Figure 2 of De Pasquale et al. 2015) does
show an early peak, which is consistent with our results.

\section{CONCLUSIONS}
Optical re-brightenings appearing at $\sim 10^4$ s in the afterglows of some GRBs
may come from a common origin. In this work, we attribute the re-brightenings to
the RS propagating into the ultrarelativistic $e^+e^-$ wind ejected by
the central engine.
We compared two methods used for solving the dynamics and found
that they are both appropriate in describing the FS-RS system.
Multi-wavelength afterglows of four GRBs can be well explained in the framework
of our scenario.

The success of our model can give helpful information on
the central engine of these GRBs. $T_{\rm sd}$ can be derived from the
X-ray peak time of the RS component as described in Section 3,
which could give clues on the characteristics of the newly born magnetar.
Generally, an earlier re-brightening means $B_{\rm NS}$ is larger
or $P_{\rm NS}$ is smaller, i.e., the magnetar losses its energy more quickly.
In the future, if the observational data can help to constrain
the parameters of the FS well enough, we will be able to
derive $B_{\rm NS}$ more accurately by using the re-brightening. It will be
a useful way to probe the characteristics of the magnetar.
Note that the derived values of $\Gamma_4$ of GRB 091029 and
GRB 100814A are larger than those of the other two GRBs.
It means that the efficiency of converting the Poynting-flux
to the kinetic energy of particles is higher in these cases.

Observationally, it seems there are no equivalent
features in X-rays around the optical re-brightenings.
However, the predicted emission near $T_{{\rm peak},2}$ in both optical and X-ray wavelengths
is dominated by the RS component in our model.
There are two main factors to result in this puzzle.
First, the actual peak time of the RS component in optical bands
is delayed ($> T_{\rm sd}$) by the EATS effect, which causes
the flux ratio $F_{{\rm opt},3}/ F_{{\rm opt},2}$ to be relatively larger since
$F_{{\rm opt},2}$ is decreasing. Second, at the X-ray peak time,
the flux ratio of $F_{{\rm X},3}/ F_{{\rm X},2}$ is expected to be relatively small.
Moreover, the emerging of the RS component in X-rays is essential
to produce the X-ray plateau, otherwise the X-ray light curve
will decay as a power-law after $T_{{\rm peak},1}$
(see $F_{\nu,2}$ in Figure 2).
In other words, the X-ray plateau should be a common feature accompanying
the optical re-brightening.

The fact that the re-brightening time is coincident with $T_{\rm sd}$,
together with the arguments that magnetars could serve
as the central engines of GRBs in recent years,
motivates us to suggest that there may be a common origin for the re-brightening.
$e^+e^-$ winds are natural outcome from magnetars, as
hinted from phenomena associated with pulsar wind nebulae. Thus we
believe our scenario could work well and it is physically reasonable.
Of course, the models involving late shells with a range of velocities
may also produce the similar results.
Further study on the form of the injected energy are thus called to give a conclusion.
In principle, the mass density is different between a baryonic shell and
an electron-positron wind. But it is not clear how the mass density can
be measured from observations. For the baryonic shell, one may calculate
the total mass of the shell to check whether it is reasonable.

In the future, two potential ways should help to discriminate between our
$e^+e^-$ winds model and models involving baryons-dominated shells.
First, the high energy emission ($>$ 100 MeV) should be more significant when
the continuously injected energy is carried by leptons rather than baryons,
since IC scattering will be much stronger \citep{Yu07b}.
As an example, Fig. 3 shows that the SSC flux from the RS exceeds
the synchrotron emission from the FS at late times in the case of GRB 100814A.
If a high energy bump is observed near the peak time of the optical re-brightening,
it will be a strong evidence for the $e^+e^-$ wind.
Furthermore, the evolution of the polarization degree during the
plateau or the re-brightening phase in two models are significantly different \citep{Lan16}.
Therefore, the observations on the high energy emission and the polarization
during the re-brightening phase would identify which model is preferred.

The $e^+e^-$ wind scenario is not necessarily based on magnetars as central engines.
A newly born BH accompanied by an accretion disk may also emit continuous
Poynting-flux \citep{Blandford77} and the luminosity evolution may somehow
be similar to that of a megnetar (Equation 1).
Such a BH can also serve as the central engine in our model.
In our model, the Poynting-flux is assumed to be converted to an $e^+e^-$ wind
efficiently, but note that the possibility of this conversion and its efficiency are still
quite uncertain for a newly born magnetar. A more general treatment is to introduce
a magnetization parameter $\sigma$ to describe the late ejecta \citep{Zhang05},
and this may affect the RS emission if $\sigma$ is significantly larger than 1.
The diversity of $\sigma$ would also help to explain why optical re-brightenings
do not appear in some GRB afterglows.
For a magnetized ejecta, the parameters derived may be correspondingly different.
However, the main conclusions of this study would remain unchanged.

\acknowledgments
We are very grateful to an anonymous referee for valuable suggestions. We thank Yiming Hu and Bing Zhang for helpful discussions.
This work was supported by the National Basic Research Program (``973" Program) of China (grant Nos. 2014CB845800 and 2013CB834900) and the National Natural Science Foundation of China (grant Nos. 11473012, 11573014 and 11322328).
X. F. Wu acknowledges support by the One-Hundred-Talents Program, the Youth Innovation Promotion Association (2011231), and the Strategic Priority Research Program ``The Emergence of Cosmological Structures'' of the Chinese Academy of Sciences (Grant No. XDB09000000). Liang Li acknowledges support by the Swedish National Space Board, and the Erasmus Mundus Joint Doctorate Program by Grant Number 2013-1471 from the EACEA of the European Commission. This work made use of data supplied by the UK Swift Science Data Center at the University of Leicester.

\appendix
\section{HYDRODYNAMICS}
\subsection{Energy Conservation Method}
Let's consider a relativistic outflow with an initial mass of $M_{\rm ej}$
propagating into a cold interstellar medium (ISM).
Two shocks separate the system into four regions:
(1) the unshocked ISM, (2) the shocked ISM, (3) the shocked wind,
and (4) the unshocked wind.
Regions 2 and 3 can be regarded as simple homogenous shells \citep{Piran99}.
In this paper, the quantities (e.g., the electron Lorentz factor $\gamma_e$, internal energy $U$,
and pressure $p$) of Region ``$i$'' are denoted by subscripts ``$i$'', and the
superscript prime ($\prime$) is used to denote the quantities in the shock
comoving frame while characters without prime denote quantities in the
observer frame.

The FS and the RS can be described by a common bulk Lorentz factor
$\Gamma_2 = \Gamma_3 = \Gamma$ (the corresponding dimensionless speed is $\beta$)
and we assume the Lorentz factor of the unshocked wind is $\Gamma_4$.
Below, $\Gamma_{ij}$ and $\beta_{ij}$
are the relative Lorentz factor and dimensionless speed of Region ``$i$''
as measured in the frame of Region ``$j$''.
Applying the jump conditions to the FS, the thermodynamical quantities of
the gases in the rest frame of Region 2 are given by:
$U_2^{\prime} = (\Gamma-1) m_2 c^2$ (internal energy) and
$p_2^{\prime} V_2^{\prime} = (\hat{\gamma_2} - 1) U_2^{\prime}$
(product of pressure and volume), where $m_2$ is the total mass
swept by the FS, $\hat{\gamma_2} \simeq (4\Gamma+1)/(3\Gamma)$ is the adiabatic index
and $c$ is the speed of light.
If the fraction of the thermal energy lost due to radiation is $\epsilon_2$
(i.e., the radiation efficiency), then the energy of Region 2 is \citep{Peer12,Geng14}
\begin{equation}
H_{2} \simeq (\Gamma-1) (m_2+M_{\rm ej}) c^2 + (1-\epsilon_2) \Gamma (U_2^{\prime} + p_2^{\prime} V_2^{\prime}).
\end{equation}
Taking $\epsilon_e$ as the equipartition parameter for
shocked electrons, the radiation efficiency can be calculated as
$\epsilon_2 = \epsilon_e t_{\rm syn}^{\prime -1}/(t_{\rm syn}^{\prime -1}+t_{\rm ex}^{\prime -1})$ \citep{Dai99},
where $t_{\rm syn}^{\prime}$ is the synchrotron cooling timescale and $t_{\rm ex}^{\prime}=R/(\Gamma c)$
is the expansion timescale in the comoving frame.
Similarly, the energy of Region 3 is
\begin{equation}
H_{3} \simeq (\Gamma-1) m_3 c^2 + (1-\epsilon_3) \Gamma (U_3^{\prime} + p_3^{\prime} V_3^{\prime}),
\end{equation}
where $U_3^{\prime} = (\Gamma_{43}-1) m_3 c ^2$ and
$p_3^{\prime} V_3^{\prime} = (\hat{\gamma_3} - 1) U_3^{\prime}$. The adiabatic index of Region 3 can be approximated as $\hat{\gamma_3}\simeq (4\Gamma_{43}+1)/3\Gamma_{43}$.
The total energy of the matter between the FS and the RS is thus
$H_{\rm tot} = H_2+H_3$.

We can derive the differential equation for the evolution of $\Gamma$
following the procedure of \cite{Huang99,Huang00} and \cite{Geng14}.
When a mass of $d m_2$ of the ISM is swept up by the FS,
a fraction of thermal energy,
$dE_{\rm loss} = \epsilon_2 \Gamma \hat{\gamma_2} (\Gamma-1) d m_2 c^2$
would be radiated from Region 2.
Similarly, when the wind matter of mass $d m_3$ is swept up by the RS,
Region 3 will gain some thermal energy, i.e.,
$dE_{\rm gain} = (1-\epsilon_3) \Gamma \hat{\gamma_3} (\Gamma_{43}-1) d m_3 c^2$.
Using $dH_{\rm tot} = dE_{\rm gain} - dE_{\rm loss}$, $d\Gamma_{43}/d\Gamma=-\beta_{43}\Gamma_{43}/\beta\Gamma$ ($\Gamma_4=$constant), and the relevant formulae above,
we have
\begin{equation}
\frac{d \Gamma}{d m_2} = \frac{f_2}{f_1} + \frac{f_3}{f_1} \frac{d m_3}{d m_2},
\end{equation}
where $f_1 = M_{\rm ej}+m_2+m_3+\frac{1}{3}(1-\epsilon_2)(8\Gamma-3)m_2
+\frac{1}{3}(1-\epsilon_3)[4\Gamma_{43}-3-\frac{1}{\Gamma_{43}}-\frac{\beta_{43}}{\beta}
(4\Gamma_{43}+\frac{1}{\Gamma_{43}})]m_3$, $f_2 = -\frac{4}{3}(\Gamma^2-1)$, and
$f_3 = -(\Gamma-1)$. On the other hand, it can be derived that (see \citealt{Geng14})
\begin{equation}
\frac{d m_3}{d m_2} = \left(\frac{\beta_4}{\beta}-1\right) \frac{\rho_4^{\prime}}{\rho_1} \Gamma_4,
\end{equation}
where $\rho_4^{\prime}$ is the comoving density of Region 4 and
$\rho_1$ is the density of the circum-burst environment.
The evolution of $\Gamma$ can thus be calculated by using Equation (A3).

\subsection{Mechanical Method}
Another method to solve the dynamics of the FS-RS system is developed by \cite{Beloborodov06}.
Here, we briefly describe this mechanical method.
In this method, the shocked Regions 2 and 3 between the FS and RS are called the ``blast'' and
the blast is assumed to move with a Lorentz factor $\Gamma$ (the corresponding dimensionless speed is $\beta$).
In this section, the subscription ``$f$'' or ``$r$'' is used to denote the quantities just behind the FS or the RS,
thus the velocities (relative to the shock front) of the postshock medium are $\beta_f$ and $\beta_r$ respectively.
Considering the conservation of energy-momentum and mass flux in  Regions 2 and 3 between
the FS ($r_f$) and the RS ($r_r$), three equations can be obtained
\begin{eqnarray}
\frac{1}{r^2 c} \frac{d}{d t} \left(r^2 \Sigma \Gamma \right) - \Gamma \left[ \rho_r \left(\beta -\beta_r \right) + \rho_f \left(\beta_f -\beta \right) \right] = 0, \\
\frac{1}{r^2 c} \frac{d}{d t} \left(r^2 H \Gamma^2 \beta \right) - \Gamma^2 \beta \left[ h_r \left(\beta -\beta_r \right) + h_f \left(\beta_f -\beta \right) \right] = p_r - p_f, \\
\frac{1}{r^2 c} \frac{d}{d t} \left(r^2 H \Gamma \right) - \Gamma \left[ h_r \left(\beta -\beta_r \right) + \rho_f \left(\beta_f -\beta \right) \right] = \frac{\Gamma}{c} \frac{d}{dt} P - \Gamma \left[ p_r \left(\beta -\beta_r \right) + p_f \left(\beta_f -\beta \right) \right],
\end{eqnarray}
where $\Sigma = \int_{r_r}^{r_f} \rho d r$, $H = \int_{r_r}^{r_f} h d r$, $P = \int_{r_r}^{r_f} p d r$, and $H - \Sigma c^2 = 4 P$,
$\rho$, $h$, $p$ are the mass density, energy density and pressure measured in the rest frame of
the FS, respectively. Other equations derived from shock jump conditions are needed
to complete the equations above (see \citealt{Uhm11,Uhm12} for details).
These equations can be solved numerically, and $\Gamma$ can be obtained together with $\Sigma$, $H$, and $P$.

\section{RADIATION}
Electrons will be accelerated by the FS and the RS after the shocks.
As usual, we assume the accelerated electrons carry a fraction $\epsilon_{e,i}$ of
the total energy and the ratio of the magnetic field energy
to the total energy is $\epsilon_{B,i}$.
In the absence of radiation loss, the energy distribution of the shocked electrons
is usually assumed to be a power-law as
$dN_{e,i}^{\prime}/d\gamma_{e,i}^{\prime} \propto \gamma_{e,i}^{\prime -p_i} (\gamma_{m,i}^{\prime} \leq
\gamma_{e,i}^{\prime} \leq \gamma_{M,i}^{\prime})$,
where $\gamma_{e,i}^{\prime}$ is the Lorentz factor of electrons of Region $i$, and $p_i$ is the spectral index.

The minimum Lorentz factor is
\begin{equation}
\gamma_{m,i}^{\prime} = \zeta_i \epsilon_{e,i} \frac{p-2}{p-1} (\hat{\Gamma}_i-1) + 1,
\end{equation}
where $\zeta_2 = m_p/m_e$ ($m_p$ and $m_e$ are the mass of protons and electrons respectively),
$\zeta_3 = 1$, $\hat{\Gamma}_2 = \Gamma_2$, and $\hat{\Gamma}_3 = \Gamma_{43}$.
The maximum Lorentz factor is
\begin{equation}
\gamma_{M,i}^{\prime} \simeq 10^8 [B_i^{\prime} (1+Y_i)]^{-1/2},
\end{equation}
where $B_i^{\prime}$ is the comoving magnetic field strength,
and $Y_i$ is the Compton parameter that is defined as the ratio
of the inverse compton (IC) power to the synchrotron power.
The Compton parameter of an electron with Lorentz factor of $\gamma_e^{\prime}$
can be determined by $Y_i(\gamma_e^{\prime}) = (-1+\sqrt{1+4\eta_{{\rm rad},i}\eta_{{\rm KN},i}\epsilon_{e,i}/\epsilon_{B,i}})/2$ \citep{Fan06a,He09},
where $\eta_{{\rm rad},i}$ is the fraction of energy that is radiated
due to synchrotron and IC radiation,
$\eta_{{\rm KN},i}$ is the fraction of synchrotron photons with energy below
the Klein-Nishina limit.

Considering the radiation loss, the actual electron distribution would be
characterized by the cooling Lorentz factor $\gamma_{c,i}^{\prime}$,
which is given by
\begin{equation}
\gamma_{c,i}^{\prime} = \frac{6 \pi m_e c (1+z)}{(1+Y_i) \sigma_{T} B_i^{\prime 2} (\Gamma_i + \sqrt{\Gamma_i^2-1})
t_{\rm obs}},
\end{equation}
where $t_{\rm obs}$ is the time measured in the observer's frame and
$\sigma_T$ is the Thomson cross section.
Then, the actual electron distribution should be given as the
following cases:
1. for $\gamma_{c,i}^{\prime}\leq\gamma_{m,i}^{\prime}$,
\begin{equation}
\frac{dN_{e,i}^{\prime}}{d\gamma_{e,i}^{\prime}}\propto\left\{
\begin{array}{ll}
\gamma_{e,i}^{\prime -2},&\gamma_{c,i}^{\prime}\leq\gamma_{e,i}^{\prime}\leq\gamma_{m,i}^{\prime},\\
\gamma_{e,i}^{\prime -p_i-1},&\gamma_{m,i}^{\prime}<\gamma_{e,i}^{\prime}\leq\gamma_{M,i}^{\prime}.
\end{array}\right.
\end{equation}
2. for $\gamma_{m,i}^{\prime}<\gamma_{c,i}^{\prime}\leq\gamma_{M,i}^{\prime}$,
\begin{equation}
\frac{dN_{e,i}^{\prime}}{d\gamma_{e,i}^{\prime}}\propto\left\{
\begin{array}{ll}
\gamma_{e,i}^{\prime -p_i},&\gamma_{m,i}^{\prime}\leq\gamma_{e,i}^{\prime}\leq\gamma_{c,i}^{\prime},\\
\gamma_{e,i}^{\prime -p_i-1},&\gamma_{c,i}^{\prime}<\gamma_{e,i}^{\prime}\leq\gamma_{M,i}^{\prime}.
\end{array}\right.
\end{equation}

With the electron distribution determined, the synchrotron emissivity of electrons
in Region $i$ at frequency $\nu^{\prime}$ can be calculated as \citep{Rybicki79}
\begin{equation}
\varepsilon_{i,\rm{syn}}^{\prime} (\nu^{\prime}) = \frac{\sqrt{3} q_e^3 B_i^{\prime}}{m_e c^2}
\int_{\min{\{\gamma_{m,i}^{\prime},\gamma_{c,i}^{\prime}\}}}^{\gamma_{M,i}^{\prime}} \frac{dN_{e,i}^{\prime}}{d\gamma_{e,i}^{\prime}}
F\left(\frac{\nu^{\prime}}{\nu_{\rm ch}^{\prime}}\right) d \gamma_{e,i}^{\prime},
\end{equation}
where $\nu_{\rm ch}^{\prime} = 3\gamma_{e,i}^{\prime 2} q_e B_i^{\prime} / (4 \pi m_e c)$,
$q_e$ is the electric charge of an electron,
$F(x) = x \int_x^{\infty} K_{5/3} (x) d x$, with $K_{5/3}(x)$ being the Bessel function.
The synchrotron self-absorption (SSA) effect should be taken into account \citep{Rybicki79,Wu03}.
In our calculation, the absorption effect can be included by multiplying
a factor to the emissivity above.
For the shell geometry, the absorption factor is expressed as $(1-e^{-\tau_{\nu^{\prime}}}) / \tau_{\nu^{\prime}}$,
where $\tau_{\nu^{\prime}}$ is the optical depth.
The self-absorption coefficients $\kappa_{\nu^{\prime}}$ can be derived analytically (see Appendix C),
which further gives
$\tau_{\nu^{\prime}} = \kappa_{\nu^{\prime}} \Delta = \frac{\kappa_{\nu^{\prime}}}{n_{e}} \frac{N_{e,{\rm tot}}}{\Omega R^2}$,
where $N_{e,{\rm tot}}$ is the total number of electrons,
$\Delta$ is the width of the shell,
$\Omega = 2 \pi (1-\cos \theta_{j})$ is the solid angle of the outflow,
and $\theta_j$ is the half-opening angle of the jet.

Besides the synchrotron radiation, electrons would be cooled by IC scattering of seed photons.
IC scattering by self-emitted synchrotron photons is referred as synchrotron self-Compton (SSC) process,
and IC scattering by photons from other regions is referred to as cross inverse-Compton (CIC) process.
If the flux density of seed photons of Region $j$ is $f_{\nu_{s,j}^{\prime}}$,
the IC ($i=j$ for SSC and $i \neq j$ for CIC) emissivity at frequency $\nu^{\prime}$ is calculated by
\citep{Blumenthal70,Yu07b}
\begin{equation}
\varepsilon_{i,\rm{IC}}^{\prime} (\nu^{\prime}) = 3 \sigma_{\rm T}
\int_{\gamma_{{\rm min},i}^{\prime}}^{\gamma_{M,i}^{\prime}} \frac{dN_{e,i}^{\prime}}{d\gamma_{e,i}^{\prime}}
d \gamma_{e,i}^{\prime} \int_{\nu_{s,j,{\rm min}}^{\prime}}^{\infty} d \nu_{s,j}^{\prime}
\frac{\nu^{\prime} f_{\nu_{s,j}}^{\prime}}{4 \gamma_{e,i}^{\prime 2} \nu_{s,j}^{\prime 2}} g(x,y),
\end{equation}
where $\gamma_{{\rm min},i}^{\prime} = \max[\min[\gamma_{c,i}^{\prime},\gamma_{m,i}^{\prime}],
h \nu^{\prime}/(m_e c^2)]$,
$\nu_{s,j,{\rm min}}^{\prime} = \nu^{\prime} m_e c^2 /
[4 \gamma_{e,i}^{\prime} (\gamma_{e,i}^{\prime} m_e c^2 - h\nu^{\prime})]$,
$x = 4 \gamma_{e,i}^{\prime} h \nu_{s,j}^{\prime} / m_e c^2$,
$y = h \nu^{\prime} / [x (\gamma_{e,i}^{\prime} m_e c^2 - h \nu^{\prime})]$,
and $g(x,y) = 2y \ln y + (1+2y)(1-y) + \frac{x^2 y^2}{2 (1+xy)} (1-y)$.

Let $\theta$ be the angle between the velocity of the emitting material
and the line of sight, the Doppler factor is then $\mathcal{D}_i = 1/[\Gamma_i (1-\beta_i \cos \theta)]$.
The observed synchrotron and IC flux densities at frequency $\nu$ ($\nu = \nu^{\prime}/\mathcal{D}_i$)
from Region $i$ are given by
\begin{equation}
F_{\nu,i}^{\rm{syn}} = \int_0^{\theta_j} d \theta V_i^{\prime}(\theta) \frac{\sin \theta}{1-\cos \theta_j}
\frac{ \mathcal{D}_i^3\varepsilon_{i,\rm{syn}}^{\prime}(\mathcal{D}_i^{-1} \nu)}{4 \pi D_L^2},
\end{equation}
\begin{equation}
F_{\nu,i}^{\rm{IC}} = \int_0^{\theta_j} d \theta V_i^{\prime}(\theta) \frac{\sin \theta}{1-\cos \theta_j}
\frac{\mathcal{D}_i^3 \varepsilon_{i,\rm{IC}}^{\prime}(\mathcal{D}_i^{-1} \nu)}{4 \pi D_L^2 },
\end{equation}
where $V_i^{\prime} (\theta)$ is the volume of the emitting material.
The luminosity distance $D_L$ is obtained by adopting a flat $\Lambda$CDM universe,
in which $H_0 = 71$~km~s$^{-1}$, $\Omega_{\rm m} = 0.27$, and $\Omega_{\Lambda} = 0.73$.
In our calculations, the integration is performed over the
equal-arrival-time surface (EATS, \citealt{Waxman97,Granot99,Huang07}).
The EATS (or a sequence of $V_i^{\prime}$) at $t_{\rm obs}$ is
determined by
\begin{equation}
t_{\rm obs} = (1+z) \int_0^{R_{\theta}} \frac{1-\beta_i \cos \theta}{\beta_i c} dr \equiv {\rm const},
\end{equation}
from which $R_{\theta}$ (or $V_i^{\prime}(\theta)$) can be derived for a given $\theta$.

\section{SYNCHROTRON SELF-ABSORPTION COEFFICIENTS}
In this section, as an example, we consider only one emitting region,
thus we remove the subscript $i$.
In the comoving frame, the synchrotron radiation power at frequency $\nu^{\prime}$
from an electron of $\gamma_e^{\prime}$ is
\begin{equation}
P^{\prime} (\nu^{\prime}) = \frac{\sqrt{3} q_e^3 B^{\prime}}{m_e c^2} \
F\left(\frac{\nu^{\prime}}{\nu_{\rm ch}^{\prime}}\right)
= C_0 F\left(\frac{\nu^{\prime}}{\nu_{\rm ch}^{\prime}}\right),
\end{equation}
where $\nu_{\rm ch}^{\prime} = 3\gamma_e^{\prime 2} q_e B^{\prime} / (4 \pi m_e c)$.
For electrons with a distribution of $dN_e^{\prime}/d\gamma_e^{\prime}$, the self-absorption coefficient at frequency $\nu^{\prime}$
can be calculated as \citep{Rybicki79}
\begin{equation}
\kappa_{\nu^{\prime}} = -\frac{1}{8 \pi m_e \nu^{\prime 2}}
\int_{\min[\gamma_{e,m}^{\prime},\gamma_{e,c}^{\prime}]}^{\gamma_{e,M}^{\prime}} \gamma_e^{\prime 2}
\frac{d}{d \gamma_e^{\prime}} \left[\frac{1}{\gamma_e^{\prime 2}} \frac{d N_e^{\prime}}{d\gamma_e^{\prime}} \right]
P^{\prime} (\nu^{\prime},\gamma_e^{\prime}) d\gamma_e^{\prime}.
\end{equation}
Define $x = \nu^{\prime}/\nu_{\rm ch}^{\prime}$, then
$\gamma_e^{\prime} = \left(\frac{4 \pi m_e c}{3 q_e B^{\prime}}\right)^{1/2}
\left(\frac{\nu^{\prime}}{x}\right)^{1/2} = C_1 \left(\frac{\nu^{\prime}}{x}\right)^{1/2}$.
We assume the distribution of electrons is
$dN_e^{\prime}/d\gamma_e^{\prime} = C_2 \gamma_e^{\prime -p}$, then
Equation (C2) can be written as
\begin{equation}
\kappa_{\nu^{\prime}} = \frac{C_0 C_1^{-p} C_2 (p+2)}{16 \pi m_e}
\left[ \int_{x_M}^{\min[x_m,x_c]} x^{(p-2)/2} F(x) d x\right] \nu^{\prime -(p+4)/2}.
\end{equation}
where $x_m = \nu^{\prime}/\nu_{m}^{\prime}$ and $x_M = \nu^{\prime}/\nu_{M}^{\prime}$.

In the slow cooling case,
\begin{equation}
\frac{dN_{e}^{\prime}}{d\gamma_{e}^{\prime}} = \left\{
\begin{array}{ll}
C_2 \gamma_{e}^{\prime -p},&\gamma_{m}^{\prime}\leq\gamma_{e}^{\prime}\leq\gamma_{c}^{\prime},\\
C_2 \gamma_c^{\prime} \gamma_{e}^{\prime -p-1},&\gamma_{c}^{\prime}<\gamma_{e}^{\prime}\leq\gamma_{M}^{\prime},
\end{array}\right.
\end{equation}
the integral in Equation (C3) can be performed for $\nu^{\prime}$ as (also see \citealt{Wu03}):
\begin{equation}
\kappa_{\nu^{\prime}} \simeq \left\{
\begin{array}{llll}
C_2 \frac{8 \pi^2}{9~2^{1/3} \Gamma(1/3)} \frac{p+2}{p+2/3} \frac{q_e}{B^{\prime}}
\gamma_m^{\prime -(p+4)} \left( \frac{\nu^{\prime}}{\nu_m^{\prime}} \right)^{-5/3},&\nu^{\prime}<\nu_m^{\prime},\\
C_2 \frac{\sqrt{3} \pi}{9} 2^{(p+2)/2}
\Gamma(\frac{p}{4}+\frac{11}{6}) \Gamma(\frac{p}{4}+\frac{1}{6}) \frac{q_e}{B^{\prime}}
\gamma_m^{\prime -(p+4)} \left( \frac{\nu^{\prime}}{\nu_m^{\prime}} \right)^{-(p+4)/2},
&\nu_m^{\prime}<\nu^{\prime}<\nu_c^{\prime},\\
C_2 \frac{\sqrt{3} \pi}{9} 2^{(p+3)/2}
\Gamma(\frac{p}{4}+\frac{25}{12}) \Gamma(\frac{p}{4}+\frac{5}{12}) \frac{q_e}{B^{\prime}}
\gamma_m^{\prime -(p+4)} \left( \frac{\nu_c^{\prime}}{\nu_m^{\prime}} \right)^{-(p+4)/2}
\left( \frac{\nu^{\prime}}{\nu_c^{\prime}} \right)^{-(p+5)/2},
&\nu_c^{\prime}<\nu^{\prime}<\nu_M^{\prime},\\
C_2 \sqrt{3} \pi \left(\frac{\pi}{2}\right)^{1/2} (p+2) \frac{q_e}{B^{\prime}}
\gamma_m^{\prime -(p+4)} \left( \frac{\nu_c^{\prime}}{\nu_m^{\prime}} \right)^{-(p+4)/2}
\left( \frac{\nu_M^{\prime}}{\nu_c^{\prime}} \right)^{-(p+5)/2}
\left( \frac{\nu^{\prime}}{\nu_M^{\prime}} \right)^{-5/2} e^{-\nu^{\prime}/\nu_M^{\prime}},
&\nu^{\prime}>\nu_M^{\prime},
\end{array}\right.
\end{equation}
where $\Gamma(p)$ is the Gamma function.
In the fast cooling case,
\begin{equation}
\frac{dN_{e}^{\prime}}{d\gamma_{e}^{\prime}} = \left\{
\begin{array}{ll}
C_2 \gamma_{e}^{\prime -2},&\gamma_{c}^{\prime}\leq\gamma_{e}^{\prime}\leq\gamma_{m}^{\prime},\\
C_2 \gamma_m^{\prime p-1} \gamma_{e}^{\prime -p-1},&\gamma_{m}^{\prime}<\gamma_{e}^{\prime}\leq\gamma_{M}^{\prime},
\end{array}\right.
\end{equation}
the integral in Equation (C3) gives:
\begin{equation}
\kappa_{\nu^{\prime}} \simeq \left\{
\begin{array}{llll}
C_2 \frac{4 \pi^2}{3~2^{1/3} \Gamma(1/3)} \frac{q_e}{B^{\prime}}
\gamma_c^{\prime -6} \left( \frac{\nu^{\prime}}{\nu_c^{\prime}} \right)^{-5/3},&\nu^{\prime}<\nu_c^{\prime},\\
C_2 \frac{4 \sqrt{3} \pi}{9} \Gamma(\frac{14}{6}) \Gamma(\frac{2}{3}) \frac{q_e}{B^{\prime}}
\gamma_c^{\prime -6} \left( \frac{\nu^{\prime}}{\nu_c^{\prime}} \right)^{-3},
&\nu_c^{\prime}<\nu^{\prime}<\nu_m^{\prime},\\
C_2 \frac{\sqrt{3} \pi}{9} 2^{(p+3)/2}
\Gamma(\frac{p}{4}+\frac{25}{12}) \Gamma(\frac{p}{4}+\frac{5}{12}) \frac{q_e}{B^{\prime}}
\gamma_c^{\prime -6} \left( \frac{\nu_m^{\prime}}{\nu_c^{\prime}} \right)^{-3}
\left( \frac{\nu^{\prime}}{\nu_m^{\prime}} \right)^{-(p+5)/2},
&\nu_m^{\prime}<\nu^{\prime}<\nu_M^{\prime},\\
C_2 \sqrt{3} \pi \left(\frac{\pi}{2}\right)^{1/2} (p+2) \frac{q_e}{B^{\prime}}
\gamma_c^{\prime -6} \left( \frac{\nu_m^{\prime}}{\nu_c^{\prime}} \right)^{-3}
\left( \frac{\nu_M^{\prime}}{\nu_m^{\prime}} \right)^{-(p+5)/2}
\left( \frac{\nu^{\prime}}{\nu_M^{\prime}} \right)^{-5/2} e^{-\nu^{\prime}/\nu_M^{\prime}},
&\nu^{\prime}>\nu_M^{\prime}.
\end{array}\right.
\end{equation}

\clearpage
\begin{deluxetable}{ccccc}
\tabletypesize{\scriptsize}
\tablewidth{0pt}
\tablecaption{A brief summary of the observed features of the four GRBs.\label{TABLE:Sample}}

\tablehead{%
        \colhead{Observational Characteristics} &
        \colhead{GRB 080413B} &
        \colhead{GRB 090426} &
        \colhead{GRB 091029} &
        \colhead{GRB 100814A}
        }
\startdata
Duration (s)                  & $8.0 \pm 1.0$   & $1.28 \pm 0.09$ & $39.2 \pm 5$    &  $174.5 \pm 9.5$  \\
$z$                           & $1.1$           & $2.609$         & $2.752$         &  $1.44$  \\
$E_{\gamma,\rm{iso}}$\tablenotemark{a} ($10^{52}$ erg)& $1.8$           & $\simeq 0.42$   & $8.3$           &  $\simeq 7$ \\
$T_{\rm peak,1}$\tablenotemark{b} (s)          & $\sim 90$       & $\sim 100$      & $400$           &  $\sim 300$ \\
$T_{\rm peak,2}$\tablenotemark{c} (s)          & $\sim 7000$     & $\sim 10^4$     & $\sim 10^4$     &  $\sim 10^5$ \\
Reference                     & 1  & 2,3  & 4  & 5
\enddata
\tablenotetext{a}{$E_{\gamma,\rm{iso}}$ is the energy released in the prompt phase.}
\tablenotetext{b}{$T_{\rm peak,1}$ is the time of the early peak in the optical light curves.}
\tablenotetext{c}{$T_{\rm peak,2}$ is the peak time of the re-brightening in the optical light curves.}
\tablecomments{Strictly speaking, $T_{\rm peak,1}$ should be the time of the early peak
appearing in the light curve. However, the peak may not be captured if the ground-based telescopes
do not respond quickly enough. In this case, we set $T_{\rm peak,1}$ to be roughly the time
of the first observational data.}
\tablerefs{(1) \citet{Filgas11}; (2) \citet{Levesque10}; (3) \citet{Nicuesa11};
(4) \citet{Filgas12}; (5) \citet{Pasquale15}.}
\end{deluxetable}

\begin{deluxetable}{ccccc}
\tabletypesize{\scriptsize}
\tablewidth{0pt}
\tablecaption{Parameters used in the fit of the afterglows of the four GRBs.\label{TABLE:Fitting1}}

\tablehead{%
        \colhead{Parameters} &
        \colhead{GRB 080413B} &
        \colhead{GRB 090426} &
        \colhead{GRB 091029} &
        \colhead{GRB 100814A}
        }
\startdata
$E_{K,{\rm iso}}$ ($10^{52}$ erg)  & 5.0     & 4.0      & 8.0      & 7.0 \\
$\Gamma_0$                                  & 400    & 300     & 200     & 150 \\
$\theta_j$ (rad)                              & 0.09   & 0.08    & 0.06    & 0.058 \\
$p_2$                                             & 2.05   & 2.5      & 2.2      & 2.1 \\
$\epsilon_{e,2}$                              & 0.045 & 0.015  & 0.007  & 0.023 \\
$\epsilon_{B,2}$                              & 0.015 & 0.001  & 0.5      & 0.05 \\
$B_{\rm NS}$ ($10^{14}$ G)           & 1.9     & 5.0      & 5.5      & 1.6 \\
$\Gamma_4$ ($10^4$)                   & 4.0     & 1.0      & 50.0    & 20.0 \\
$p_3$                                             & 2.8     & 2.5      & 2.3      & 2.45 \\
$\epsilon_{B,3}$                              & 0.2     & 0.001  & 0.3     & 0.18 \\
$n_1$ (cm$^{-3}$)                         & 0.1     & 50.0    & 0.02    & 0.1 \\
$\chi^2/{\rm d.o.f.}$                       & 7.5     & 4.9     & 11.6    & 20.7
\enddata
\tablecomments{~
The relatively high $\chi^2/{\rm d.o.f.}$ of GRB 100814A is due to the bad fitting to the data of g-band.
This is the shortness of our model since our model is only good at interpreting the data of which the
spectrum is strictly a broken-power-law. When we apply our model to the data with one or two bands
deviating from the power-law determined by the data of the rest bands, we would get high $\chi^2/{\rm d.o.f.}$.}
\end{deluxetable}

\begin{deluxetable}{ccccc}
\tabletypesize{\scriptsize}
\tablewidth{0pt}
\tablecaption{Parameters constrained from the fit of the afterglows of the four GRBs.\label{TABLE:Fitting2}}

\tablehead{%
        \colhead{Parameters} &
        \colhead{GRB 080413B} &
        \colhead{GRB 090426} &
        \colhead{GRB 091029} &
        \colhead{GRB 100814A}
        }
\startdata
$B_{\rm NS}$ ($10^{14}$ G) & $1.73^{+0.03}_{-0.03}$   & $5.29^{+0.14}_{-0.16}$ & $5.81^{+0.14}_{-0.04}$  &  $1.44^{+0.04}_{-0.01}$ \\
$\Gamma_4$ ($10^4$)         & $3.91^{+0.91}_{-0.61}$  & $1.07^{+0.61}_{-0.33}$  & $81.5^{+7.1}_{-26.7}$   & $21.4^{+2.8}_{-1.7}$  \\
$\epsilon_{B,3}$   & $0.197^{+0.066}_{-0.052}$ & $9.77^{+2.35}_{-2.15} \times 10^{-4}$ & $0.35^{+0.07}_{-0.05}$  & $0.22^{+0.03}_{-0.03}$ \\
\enddata
\tablecomments{~Limited by the computer resources, we performed the fitting
by setting only three parameters as free parameters, but with other parameters
fixed as constants taken from Table 2.
These three parameters, i.e., $B_{\rm NS}$, $\Gamma_4$ and $\epsilon_{B,3}$,
are more interesting since they are directly related with the re-brightening.}
\end{deluxetable}

\clearpage
\begin{figure}
   \begin{center}
   \includegraphics[scale=0.6]{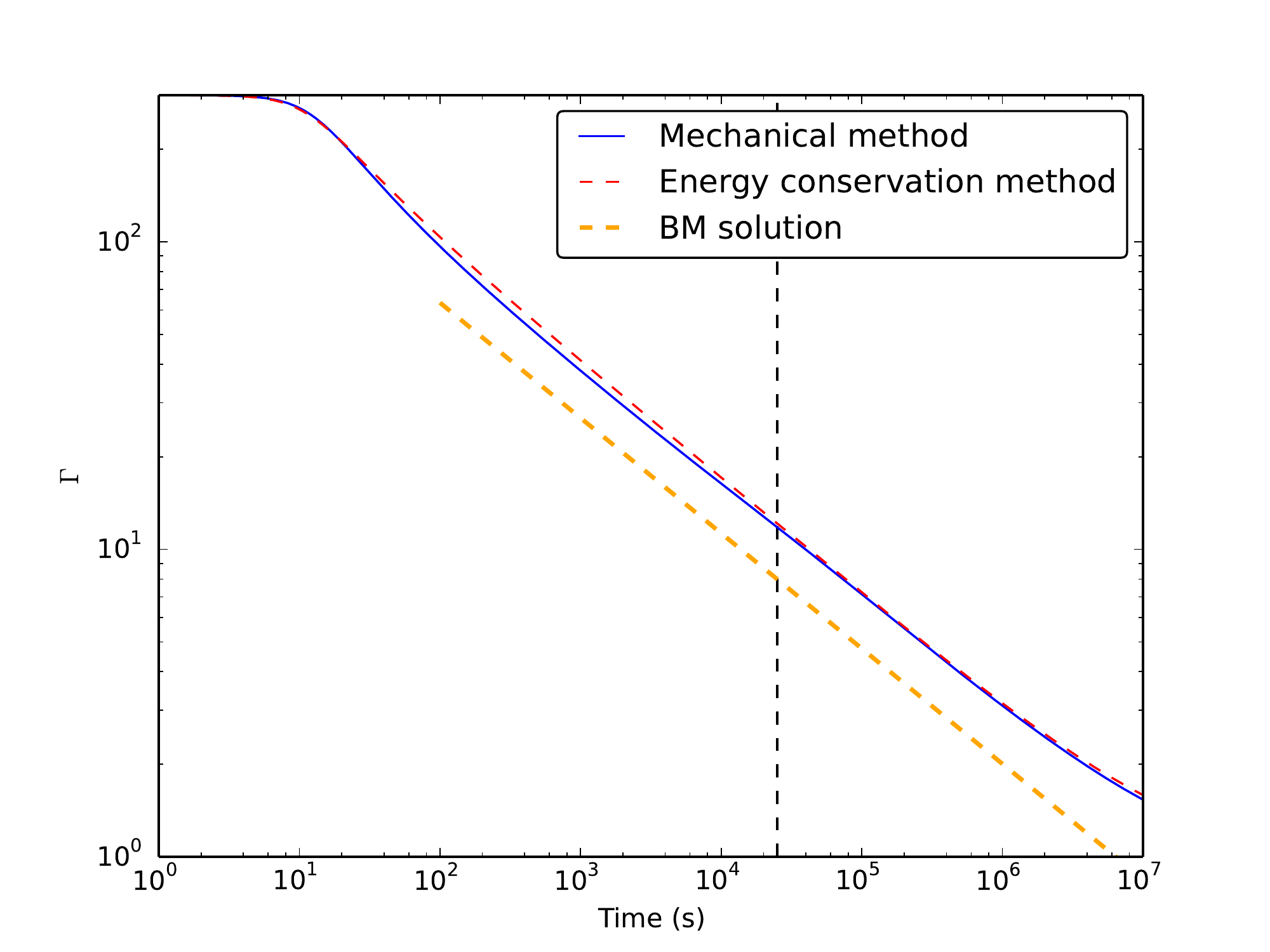}
   \caption{Temporal evolution of the bulk Lorentz factor $\Gamma$.
   The blue solid line is the result from the mechanical method and the red dashed line is
   the result from the energy conservation method. In both cases, the initial
   parameter values are: $E_{K,\rm{iso}} = 1.0 \times 10^{53}$ erg, $\Gamma_0 = 300$,
   $n_1 =1$~cm$^{-3}$, $\Gamma_4 = 10^4$, and $B_{\rm NS} = 2 \times 10^{14}$ G.
   The thick dashed orange line represents the BM solution (schematic),
   i.e., $\Gamma \propto t_{\rm obs}^{-3/8}$
   and the vertical dashed line denotes the time of $T_{\rm sd}$.}
   \label{Fig:plot1}
   \end{center}
\end{figure}

\begin{figure}
   \begin{center}
   \includegraphics[scale=0.6]{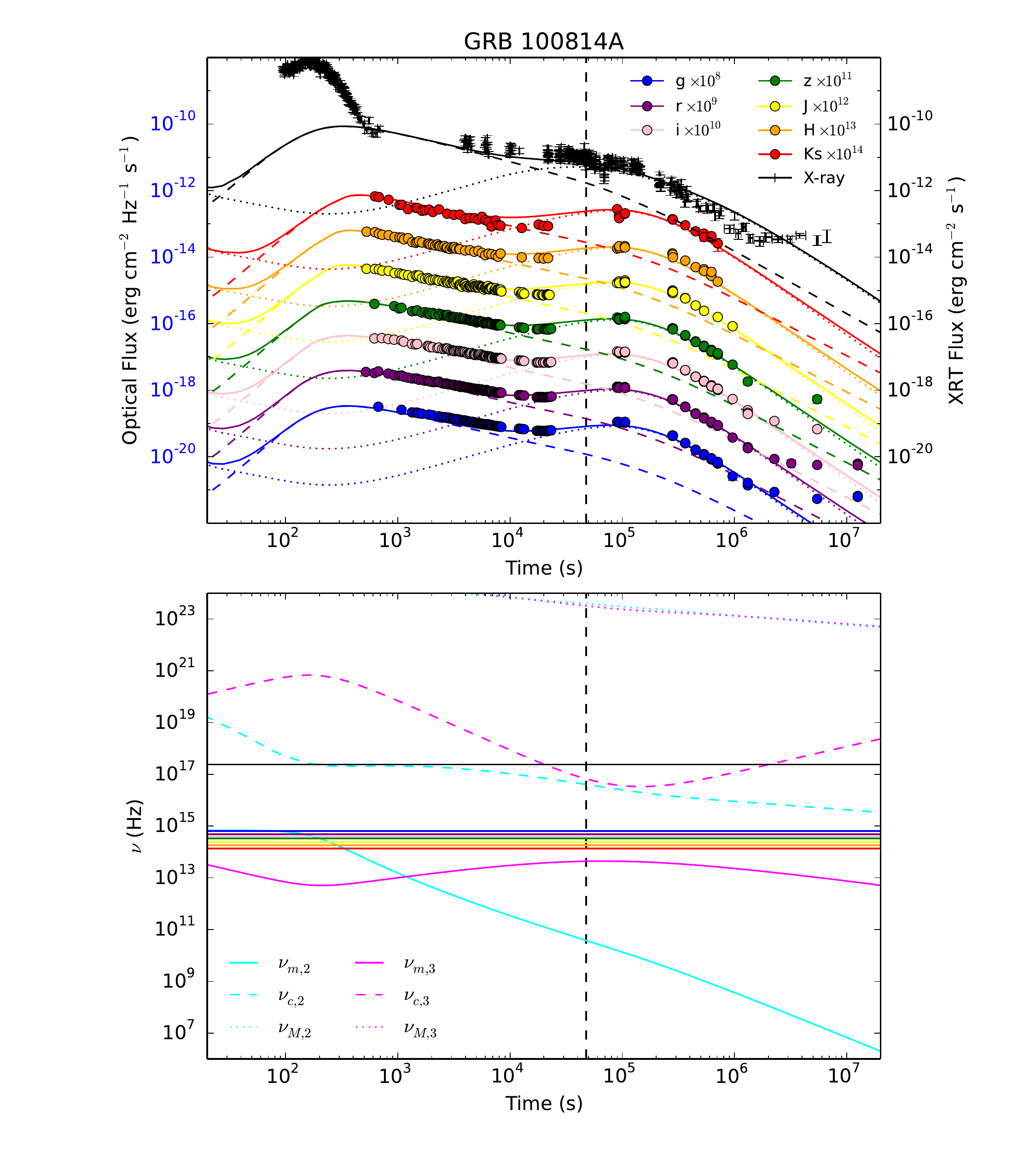}
   \caption{Fitting to the multi-wavelength afterglow of GRB 100814A by using our
   $e^+e^-$ wind injection model. The upper panel shows the fitting results in seven
   optical--infrared bands (fluxes at different bands have been multiplied by different factors for clarity).
   The optical-infrared observational data are taken from \cite{Nardini14},
   and the X-ray (0.3--10 keV) data are taken from the {\it Swift}/XRT website\protect\footnotemark.
   The dashed and dotted lines are emissions from the FS (Region 2) and the RS (Region 3)
   respectively. The solid lines are the total flux.
   The vertical dashed line denotes the time of $T_{\rm sd}$.
   The lower panel shows the evolution of $\nu_m$ (solid lines), $\nu_c$ (dashed lines),
   and $\nu_M$ (dotted lines) in Region 2 (cyan color) and Region 3 (magenta color).
   The horizontal lines represent the effective frequencies of seven optical--infrared bands
   and the X-ray band (1 keV) respectively.}
   \label{Fig:plot2}
   \end{center}
\end{figure}
\footnotetext{http://www.swift.ac.uk/xrt\_curves/00431605/}

\begin{figure}
   \begin{center}
   \includegraphics[scale=0.6]{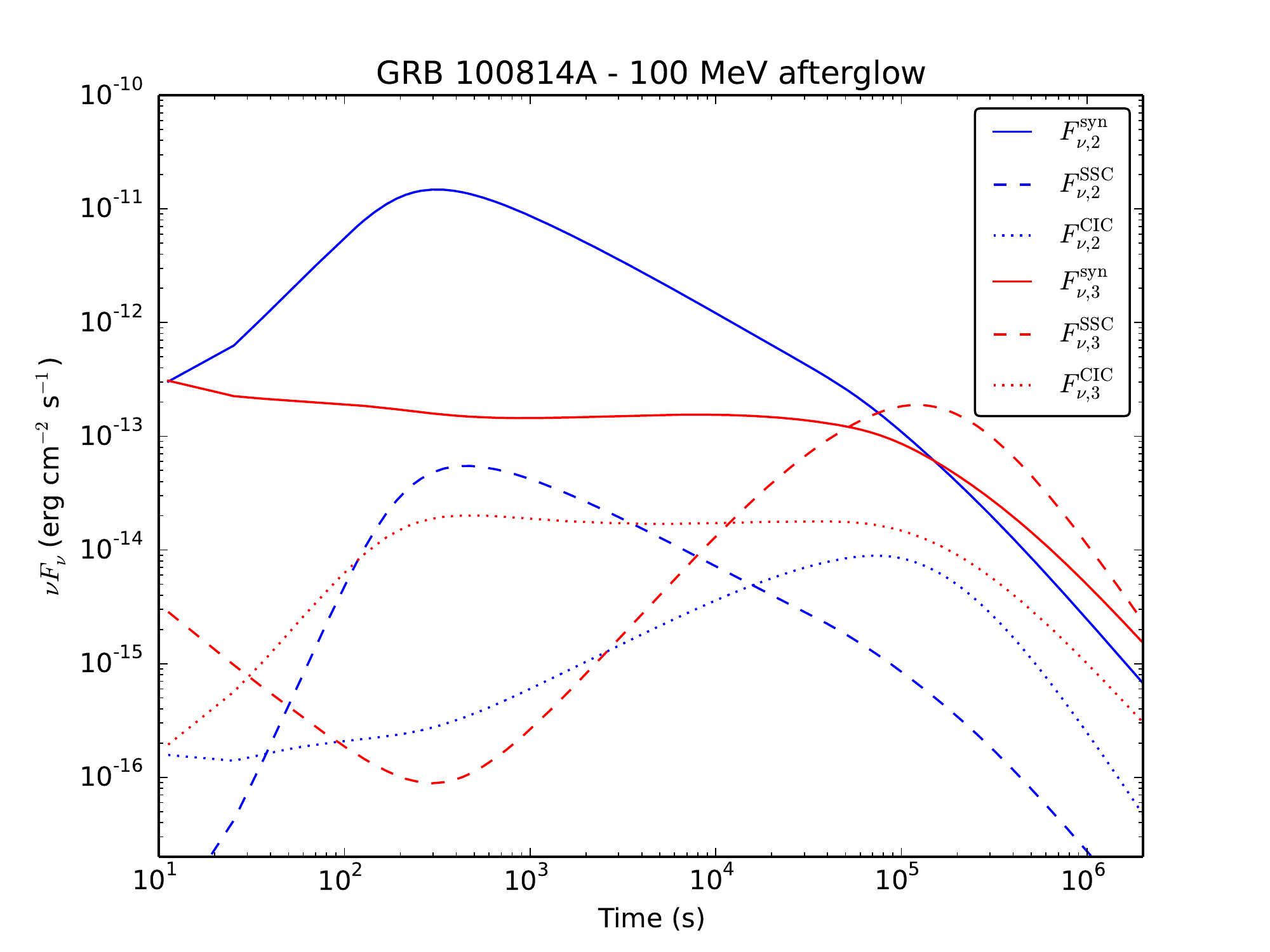}
   \caption{High energy emission (at 100 MeV) of GRB 100814A calculated from our model.
   The parameter values are the same as those in Figure 2.
   The solid lines represent the synchrotron flux, the dashed lines represent
   the SSC flux and the dotted lines represent the CIC flux.
   The blue color denotes the flux from the FS and the red color denotes the flux from the RS.
   }
   \label{Fig:plot3}
   \end{center}
\end{figure}

\begin{figure}
   \begin{center}
   \includegraphics[scale=0.6]{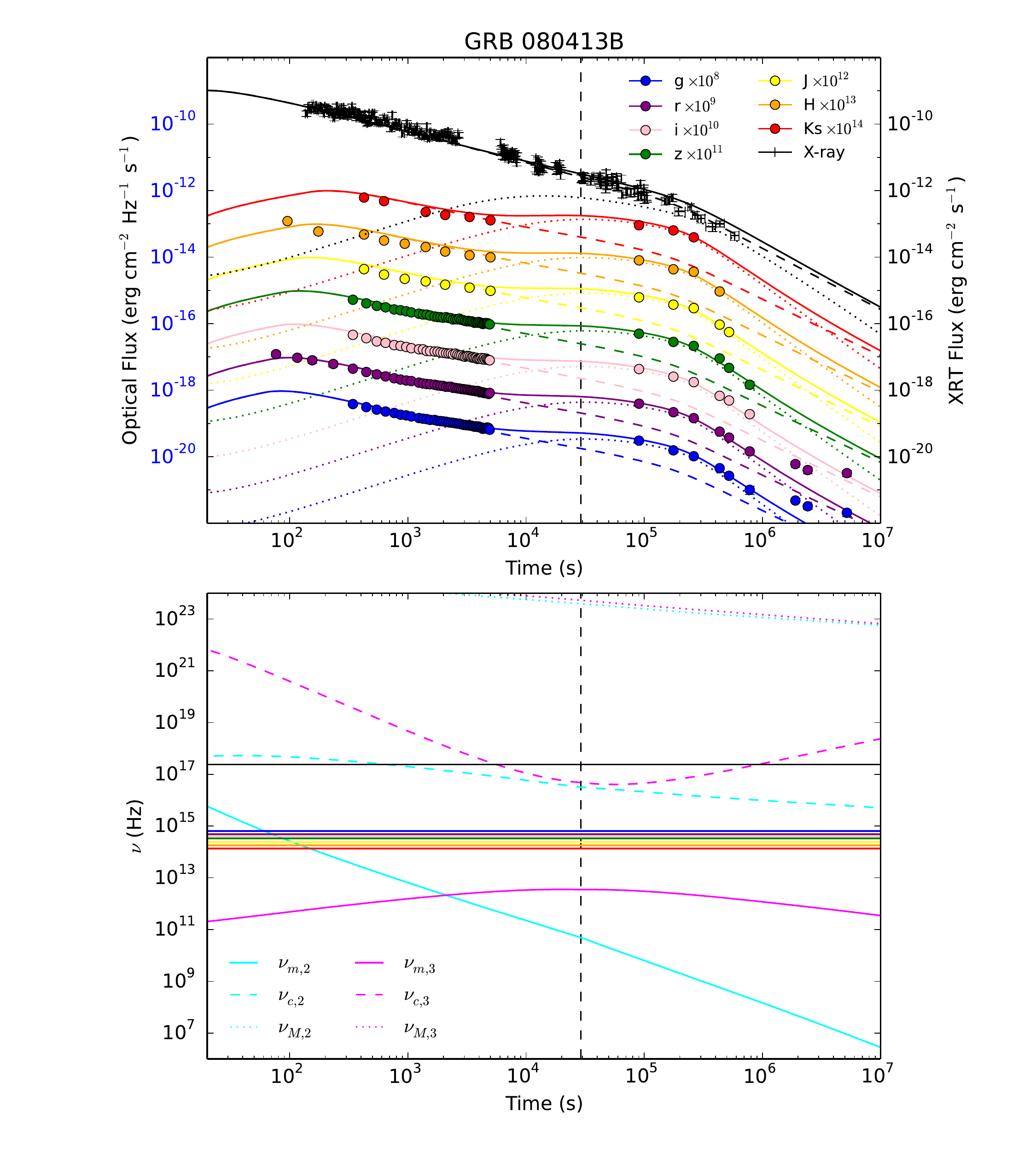}
   \caption{Fitting to the multi-wavelength afterglow of GRB 080413B by using our
   $e^+e^-$ wind injection model. In the upper panel, the optical--infrared observational data are taken from \cite{Filgas11},
   and the X-ray (0.3--10 keV) data are taken from the {\it Swift}/XRT website\protect\footnotemark.
   The dashed and dotted lines are emissions from the FS and the RS
   respectively, and the solid lines are the total flux.
   Detailed fitting parameters are listed in Table 2.
   Similar to Figure 2, the lower panel shows the evolution of some characteristic frequencies.
   }
   \label{Fig:plot4}
   \end{center}
\end{figure}
\footnotetext{http://www.swift.ac.uk/xrt\_curves/309111}

\begin{figure}
   \begin{center}
   \includegraphics[scale=0.6]{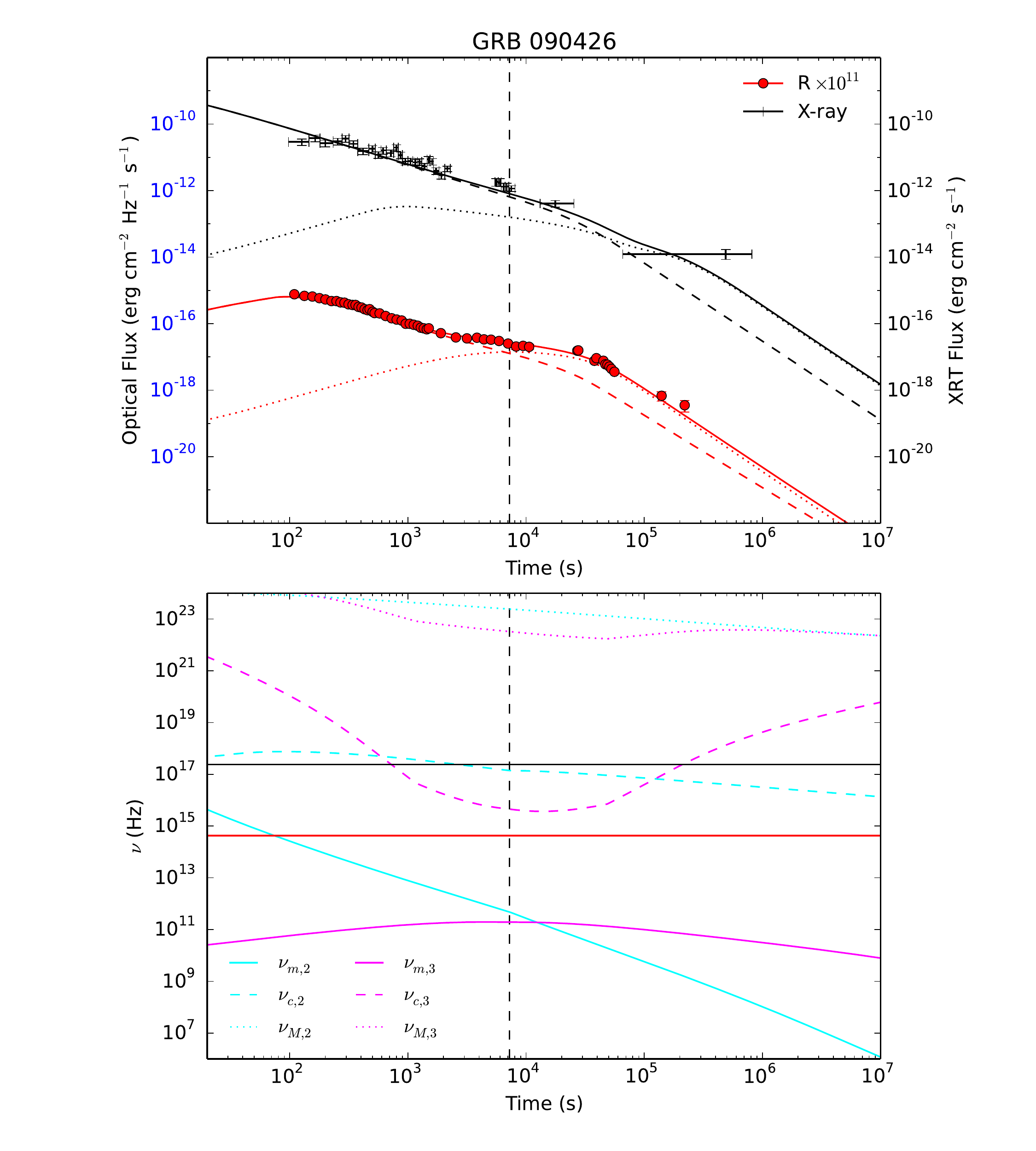}
   \caption{Fitting to the multi-wavelength afterglow of GRB 090426 by using our
   $e^+e^-$ wind injection model. In the upper panel, the observational data in R band are taken from \cite{Nicuesa11},
   and the X-ray (0.3--10 keV) data are taken from the {\it Swift}/XRT website\protect\footnotemark.
   The dashed and dotted lines are emissions from the FS and the RS
   respectively, and the solid lines are the total flux.
   Detailed fitting parameters are listed in Table 2.
   Similar to Figure 2, the lower panel shows the evolution of some characteristic frequencies.
   }
   \label{Fig:plot5}
   \end{center}
\end{figure}
\footnotetext{http://www.swift.ac.uk/xrt\_curves/350479}

\begin{figure}
   \begin{center}
   \includegraphics[scale=0.6]{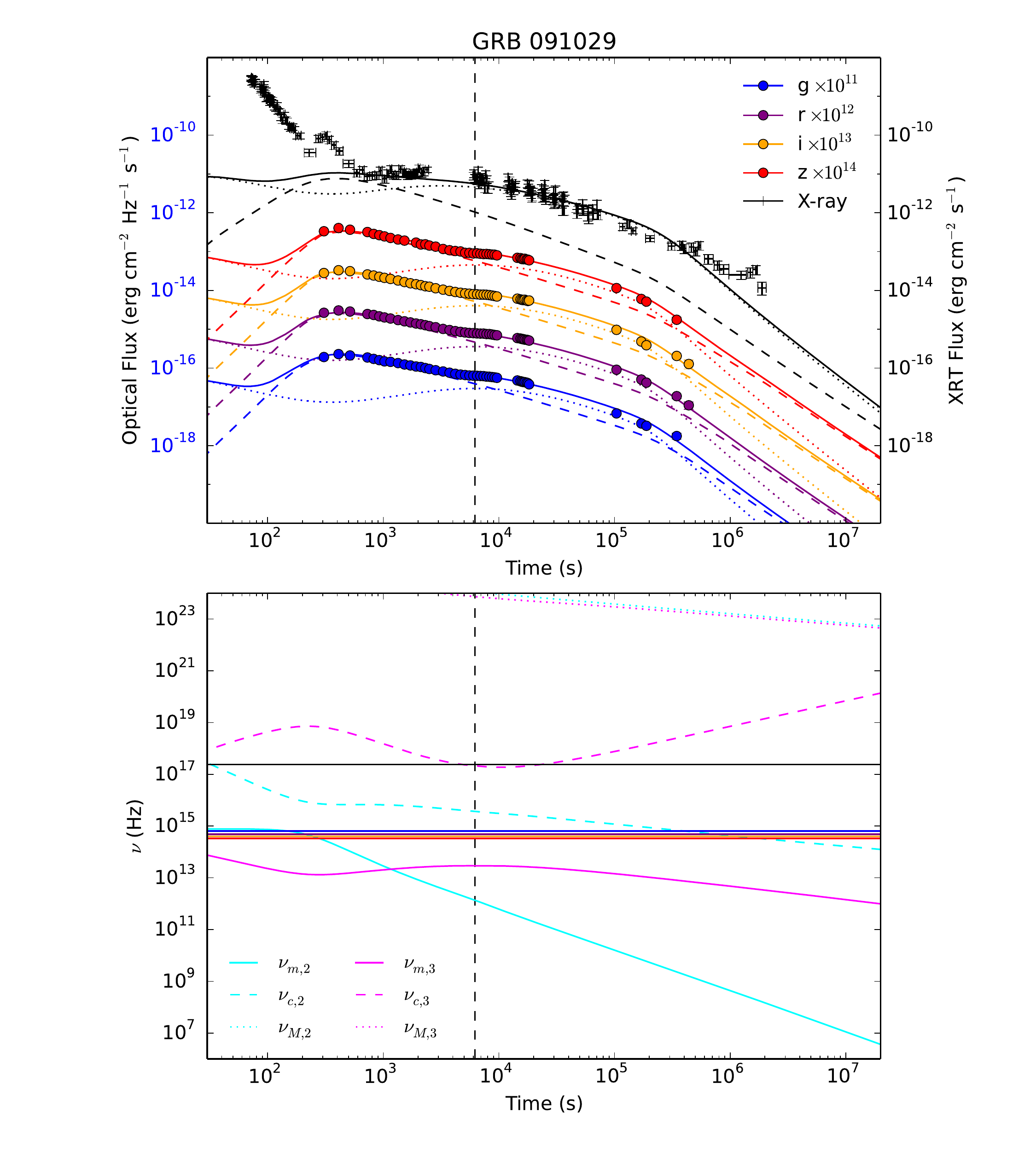}
   \caption{Fitting to the multi-wavelength afterglow of GRB 091029 by using our
   $e^+e^-$ wind injection model. In the upper panel, the optical--infrared observational data are taken from \cite{Filgas12},
   and the X-ray (0.3--10 keV) data are taken from the {\it Swift}/XRT website\protect\footnotemark.
   The dashed and dotted lines are emissions from the FS and the RS
   respectively, and the solid lines are the total flux.
   Detailed fitting parameters are listed in Table 2.
   Similar to Figure 2, the lower panel shows the evolution of some characteristic frequencies.
   }
   \label{Fig:plot6}
   \end{center}
\end{figure}
\footnotetext{http://www.swift.ac.uk/xrt\_curves/374210}

\begin{figure}
   \begin{center}
   \includegraphics[scale=0.45]{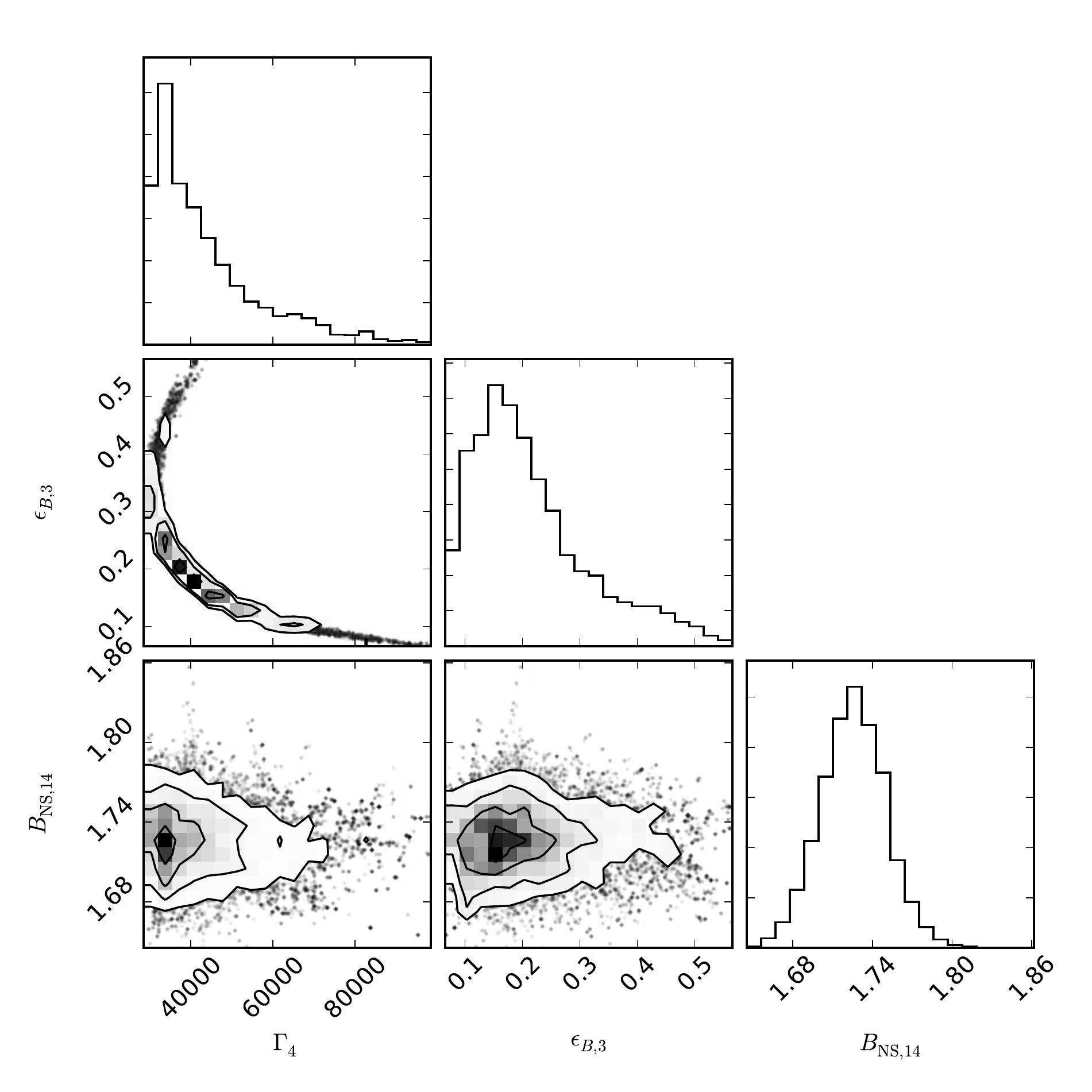}
   \caption{The MCMC fitting results for GRB 080413B.
   The marginalized distribution for each parameter is shown in the histograms along the diagonal.
   The off-diagonal contour plots show the covariances between various pairs of parameters.
   Contours are shown at 0.5, 1, 1.5, and 2 sigma.
   The best-fit values and error bars for parameters in Table 3 are calculated as the median and
   the 68\% credible intervals about the median of the marginalized posterior distribution.
   }
   \label{Fig:plot7}
   \end{center}
\end{figure}

\begin{figure}
   \begin{center}
   \includegraphics[scale=0.45]{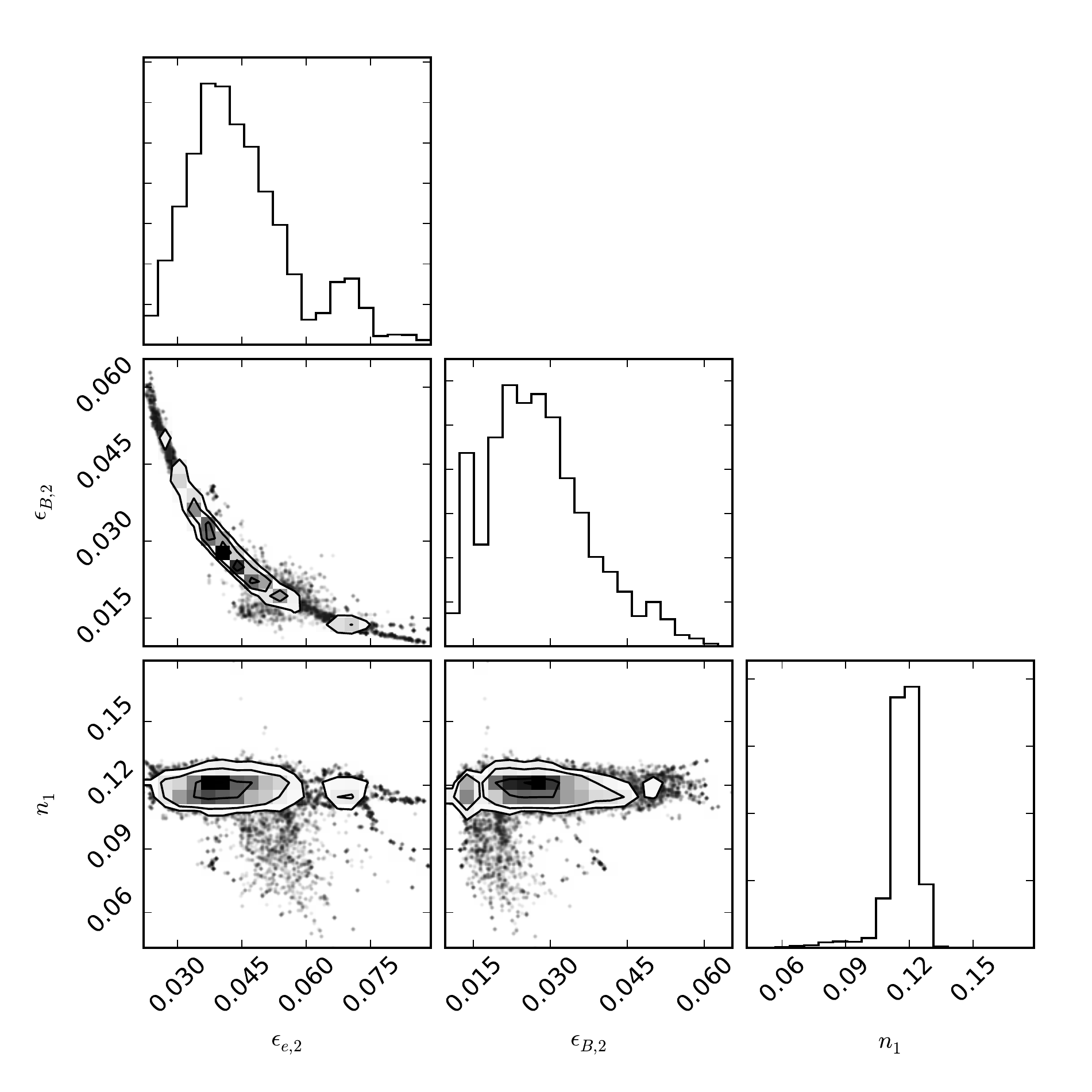}
   \caption{Similar to Figure 7 but the free parameters are $\epsilon_{e,2}$,
   $\epsilon_{B,2}$, $n_1$.
   }
   \label{Fig:plot8}
   \end{center}
\end{figure}

\end{document}